\documentclass[%
 reprint,longbibliography,preprintnumbers,
nofootinbib,
 amsmath,amssymb,
 aps
]{revtex4-1}
\pdfoutput=1
\usepackage{graphicx}
\usepackage[utf8]{inputenc}
\usepackage{flushend}
\usepackage{dcolumn}
\usepackage{bm}
\usepackage{balance}

\usepackage[normalem]{ulem}

\usepackage[colorlinks = true,
            linkcolor = blue,
            urlcolor  = blue,
            citecolor =green,
            anchorcolor = blue]{hyperref}
\usepackage{verbatim}
\usepackage{color,ulem}
\usepackage[english]{babel}

\usepackage[utf8]{inputenc}
\input Starburst.fd
\newcommand*\initfamily{\usefont{U}{Starburst}{xl}{n}}\initfamily

\newcommand{\beq}{\begin{eqnarray}}
\newcommand{\eeq}{\end{eqnarray}}
\usepackage{amsmath}
\usepackage{tikz}
\usetikzlibrary{decorations.pathmorphing}
\usetikzlibrary{shapes.misc}
\tikzset{cross/.style={cross out, draw=black, minimum size=8*(#1-\pgflinewidth), inner sep=0pt, outer sep=0pt},
cross/.default={1pt}}
\usetikzlibrary{patterns,math}
\begin{document}

\title{\huge  Plasticity in amorphous solids is mediated\\ by topological defects in the displacement field}

\author{\textbf{Matteo Baggioli}$^{1,2}$}%
 \email{b.matteo@sjtu.edu.cn}
\author{\textbf{Ivan Kriuchevskyi}$^{3}$}%
 \email{kruchevskyivan@gmail.com}
\author{\textbf{Timothy W. Sirk}$^{4}$}
\email{timothy.w.sirk.civ@mail.mil}
\author{\textbf{Alessio Zaccone}$^{3,5}$}%
 \email{alessio.zaccone@unimi.it}
 
 \vspace{1cm}
 
\affiliation{$^{1}$Wilczek Quantum Center, School of Physics and Astronomy, Shanghai Jiao Tong University, Shanghai 200240, China}
\affiliation{$^{2}$Shanghai Research Center for Quantum Sciences, Shanghai 201315.}
\affiliation{$^{3}$Department of Physics ``A. Pontremoli'', University of Milan, via Celoria 16,
20133 Milan, Italy.}
\affiliation{$^{4}$Polymers Branch, US Army Research Laboratory, Aberdeen Proving Ground, MD 21005, USA}
\affiliation{$^{5}$Cavendish Laboratory, University of Cambridge, JJ Thomson
Avenue, CB30HE Cambridge, U.K.}

\begin{abstract}
The microscopic mechanism by which amorphous solids yield plastically under an externally applied stress or deformation has remained elusive in spite of enormous research activity in recent years. Most approaches have attempted to identify atomic-scale structural ``defects'' or spatio-temporal correlations in the undeformed glass that may trigger plastic instability. In contrast, here we show that the topological defects which correlate with plastic instability can be identified, not in the static structure of the glass, but rather in the nonaffine displacement field under deformation. These dislocation-like topological defects (DTDs) can be quantitatively characterized in terms of Burgers circuits (and the resulting Burgers vectors) which are constructed on the microscopic nonaffine displacement field. We demonstrate that (i) DTDs are the manifestation of incompatibility of deformation in glasses as a result of violation of Cauchy-Born rules (nonaffinity); (ii) the resulting average Burgers vector displays peaks in correspondence of major plastic events, including a spectacular non-local peak at the yielding transition, which results from self-organization into shear bands due to the attractive interaction between anti-parallel DTDs; (iii) application of Schmid's law to the DTDs leads to prediction of shear bands at 45 degrees for uniaxial deformations, as widely observed in experiments and simulations.

\end{abstract}

\maketitle

Identifying the mechanism of plastic deformation in amorphous solids, such as glasses, is one of the major unsolved problems in condensed matter physics. In crystals, plastic flow is mediated by dislocations. These are topological defects corresponding to one missing crystalline plane in the lattice (edge dislocations) or to a lattice plane shifted by one layer (screw dislocations). While the mechanism of dislocation-mediated plastic deformation in crystals was already figured out in seminal work by Taylor~\cite{Taylor}, Polanyi~\cite{Polanyi}, and Orowan~\cite{Orowan} in 1934, a comparable mechanistic understanding of plastic deformation in glasses is still missing.

Numerical simulation studies and earlier theories of plastic activity in glasses have established the existence of so-called Shear Transformation Zones (STZs)~\cite{Falk}. These arise in regions where atomic motions are strongly nonaffine, i.e. with additional (nonaffine) displacements on top of those (affine) dictated by the macroscopic strain, that are required from mechanical equilibrium~\cite{Lemaitre,Zaccone2011}.
However, STZs have remained poorly characterized in terms of their structure and topology, until pioneering work by Procaccia and co-workers~\cite{Procaccia} suggested that STZs can be identified with Eshelby-like quadrupolar events in the displacement field that self-organize into 45-degrees shear bands to minimize the elastic energy~\cite{Procaccia} (see also~\cite{DeGiuli}). Although this mechanism of self-organization of quadrupoles can explain observations of sinusoidal density fluctuations in shear bands of metallic glasses~\cite{Wilde,Sopu}, the quadrupoles are not the only shape of plastic instabilities, and in certain systems are rarely observed or not observed at all~\cite{Rosso,Zapperi}.

In this paper, we provide the more general answer to the problem of identifying the mechanism of plastic instability in amorphous solids, and its topological nature. We start by showing that the (nonaffine) displacement field of glasses presents well defined topological singularities connected with the breakdown of the compatible deformation condition, that we demonstrate here for the first time for glasses. These topological structures are similar to dislocations (and/or vortices in superfluids), with the important difference that dislocations in crystals appear in the undeformed lattice, whereas here they appear in the displacement field under deformation. This is linked to the intrinsic out-of-equilibrium nature of glasses and it is also a fundamental difference with respect to earlier works that aimed at describing dislocations in the static structure of undeformed glass~\cite{Steinhardt1979,Steinhardt1981,Widom,Moshe}.

We show that these dislocation-like topological defects (DTDs) are the carriers of plasticity, since they lead to an average Burgers vector that strongly correlates with plastic events, and displays a strong global peak at the yielding point. This yielding peak is highly correlated throughout the material as expected for a sample-spanning slip system. Based on this evidence, a consistent theoretical description of plasticity in amorphous solids can be formulated, with predictions in excellent agreement with observations.\\

The mechanical deformation in a material can be characterized by the displacement (vector) field $u_i$ \cite{chaikin2000principles,Landau_elasticity}, which defines the deviations of the material points from their original positions ($x_i$) in the undeformed frame:
\begin{equation}
    x'_i\,=\,x_i\,+\,u_i\,.
\end{equation}
The $i$ index here indicates the different spatial directions $i=(x,y,z)$.
The displacement vector can be decomposed into its affine and nonaffine contributions \cite{PhysRevE.72.066619}
\begin{equation}
  u_i\,=\,u^{\text{A}}_i\,+\,u^{\text{NA}}_i\,=\,\Lambda^k_i\,x_k\,+\,u^{\text{NA}}_i  
\label{displacement}
\end{equation}
where ${\Lambda^k}_i$ is a matrix of constants. 
Non-zero nonaffine displacements $u^{\text{NA}}_i$ arise in glasses and non-centrosymmetric crystals in order to preserve mechanical equilibrium in the affine position dictated by the applied strain field~\cite{Lemaitre,Zaccone2011,Cui}.
In ordered crystals, the strain tensor $\epsilon_{ij}\equiv \partial _{(i}u_{j)}$ obeys the so-called \textit{compatibility constraint} \cite{M.1906,beltrami1886sull}:
\begin{equation}
    \nabla\,\times\,\nabla\,\times\,\mathbf{\epsilon}\,=\,0\,,
\end{equation}
which is equivalent to saying that $du_i$ is a closed differential form.\\
More in general, considering the total displacement field, one can define a Burgers vector \cite{kleinert1989gauge} as the line integral of the vector field $du_i$ on a closed loop $\mathcal{L}$,
\begin{equation}
    b_i\,\equiv\,-\,\oint_\mathcal{L}\,du_i\,=\,-
    \,\oint_\mathcal{L}\,\frac{du_i}{dx^k}\,dx^k\,.
    \label{Burgers}
\end{equation}
As shown below, the Burgers vector vanishes for affine displacements and it is finite for nonaffine ones:
\begin{equation}
b_i^{\text{A}}\,=\,0\,,\qquad  b_i^{\text{NA}}\,\neq\,0\,.
\end{equation}

A non-vanishing Burgers vector indicates the presence of topological defects inside the loop $\mathcal{L}$. In particular, it is associated to a non-trivial winding number around the line defect. The presence of a finite Burgers vector is equivalent to the explicit breaking of an emergent topological symmetry expressed in terms of the conservation of a two-form current $J^{\mu\nu}_I$ ~\cite{Grozdanov:2017kyl,baggioli2021deformations}:
\begin{equation}
    \partial_\mu J^{\mu\nu}_I\,\neq\,0,\,\qquad \text{with}\qquad J^{\mu\nu}_I\equiv \epsilon^{\mu\nu\rho}\partial_\rho u_I\,,
    \label{Bianchi}
\end{equation}
which plays the exact same role of the Bianchi identity in the classical covariant Maxwell formulation of electromagnetism (EM) \cite{Nussinov1,Grozdanov:2016tdf}. In other words, the presence of defects and a finite Burgers vector is in 1-to-1 correspondence to the existence of magnetic monopoles in EM \cite{Grozdanov:2018ewh}.\\
Other typical examples are those of dislocations in crystals and vortices in superfluids \cite{kleinert1989gauge,BEEKMAN20171,doi:10.1080/14786430600636328,2009arXiv0905.2996C,Ronhovde2011}. The role of these generalized global symmetries has been recently recognized to be crucial to classify topological phases of matter \'a la Landau \cite{Kapustin:2013uxa,Gaiotto:2014kfa,PhysRevB.93.155131}. For more details regarding the connection between generalized global symmetries and nonaffine displacements see the companion paper \cite{baggioli2021deformations}. 

\begin{figure}[ht]
    \centering
    \includegraphics[width=0.75\linewidth]{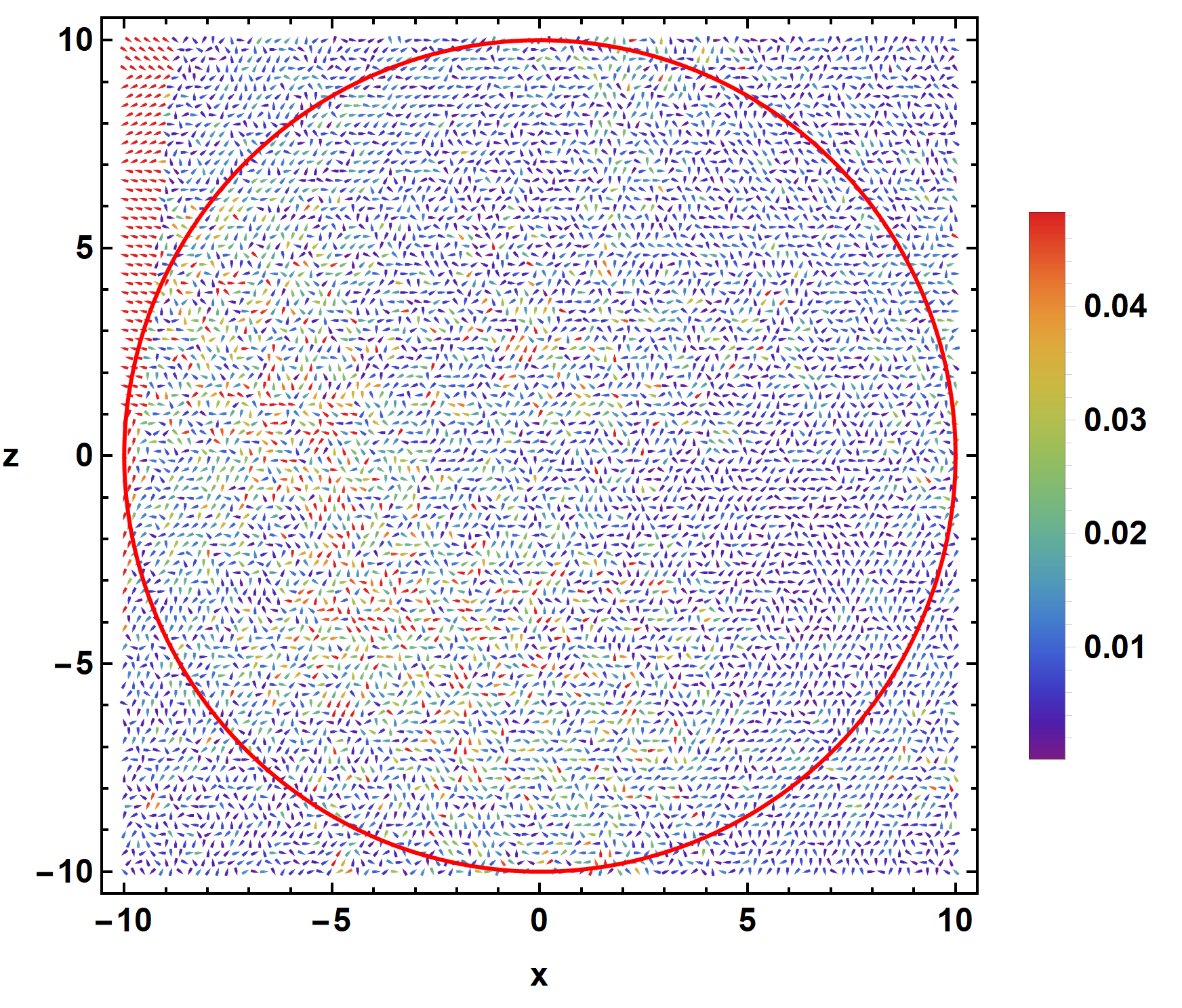}
    
    \vspace{0.3cm}
    
        \includegraphics[width=0.8\linewidth]{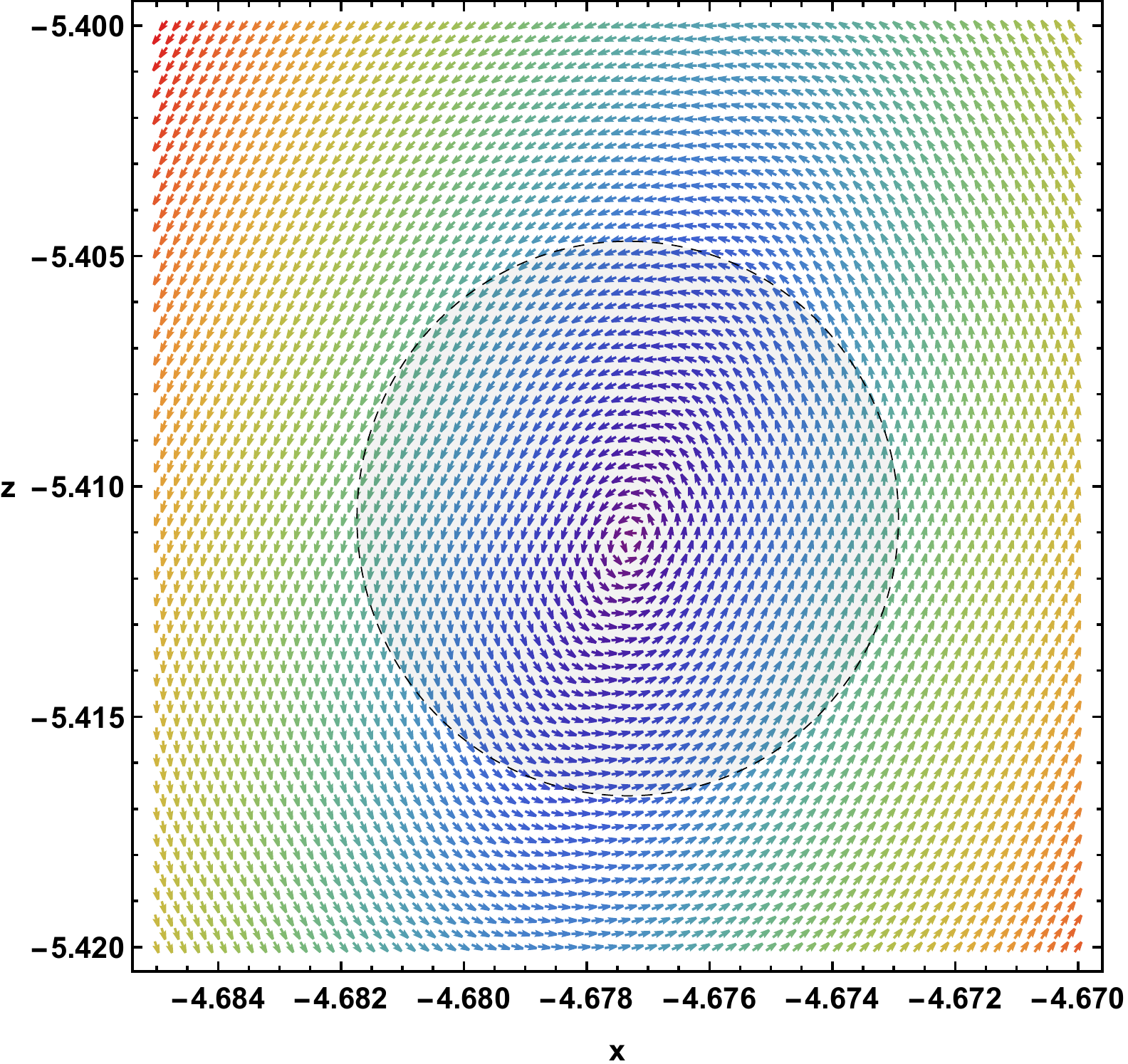}
    \caption{\textbf{Top: }A snapshot of the interpolated 2D displacement field $u_i$ for a single replica at $\gamma = 0.08$. The colors indicate the amplitude of the displacement field $|\Vec{u}|$. The red curve is the closed Burgers loop with $R=10$ on which the Burgers vector is computed using Eq.\eqref{Burgers}. \textbf{Bottom: }A zoom around a strongly nonaffine region with vortex-like shape.}
    \label{fig:main1}
\end{figure}

There \cite{baggioli2021deformations}, we showed more formally that the nonaffine dynamics typical of liquids and amorphous systems necessarily implies the presence of finite Burgers vectors and topological defects. Here, we make one step forward and we demonstrate these concepts on glass deformation data taken from numerical simulations of a coarse-grained (flexible-chain) polymer glass well below the glass transition used in previous work~\cite{Palyulin}, undergoing athermal quasistatic (AQS) shear deformation.\\

In Fig.~\ref{fig:main1} we show a typical snapshot of the displacement field at strain $\gamma=0.08$, with a system-spanning Burgers circuit. Several regions with strongly nonaffine configurations exhibiting vortex-like shape are found. At those points, the displacement field is not single valued and the integral of the Burgers vector around those region is non-zero.\\

The displacement field was measured from the MD simulation and subsequently subjected to an interpolation procedure in order to obtain a smooth field for further formal calculations (for details see the Supplementary Material~\cite{Supplementary} which includes Refs.\cite{Kremer1986,PhysRevB.102.024108,LAMMPS}). Evaluating the Burgers integral according to Eq.\eqref{Burgers} gives a non-zero Burgers vector $b_{i}$. As shown in ~\cite{Supplementary}, the same calculation on a purely affine displacement field, gives $b_{i}=0$.
Then in Eq.\eqref{Bianchi}, this implies that the displacement field is single-valued and $\partial_\mu J^{\mu\nu}_I\neq 0$. This also implies the violation of the compatibility condition~\cite{Bassani} already in the small deformation (elastic) regime of glasses, which was speculated to occur when the deformation is nonaffine~\cite{Zimmerman}, and that we demonstrate here for the first time for glasses. This finding also indicates that not only the reference metric space is curved~\cite{Moshe}, but also that the affine connections (Christoffel symbols) are not symmetric in their lower indices and the Einstein-Cartan torsion tensor is non-zero~\cite{Tartaglia}. Importantly, while the above facts have been established in crystal plasticity for large plastic deformations~\cite{Bassani}, we demonstrate here microscopically that they apply to glasses even in the elastic infinitesimal deformation regime, providing a direct link between geometry and plasticity.\\

\begin{figure}[ht]
    \centering
    \includegraphics[width=0.8\linewidth]{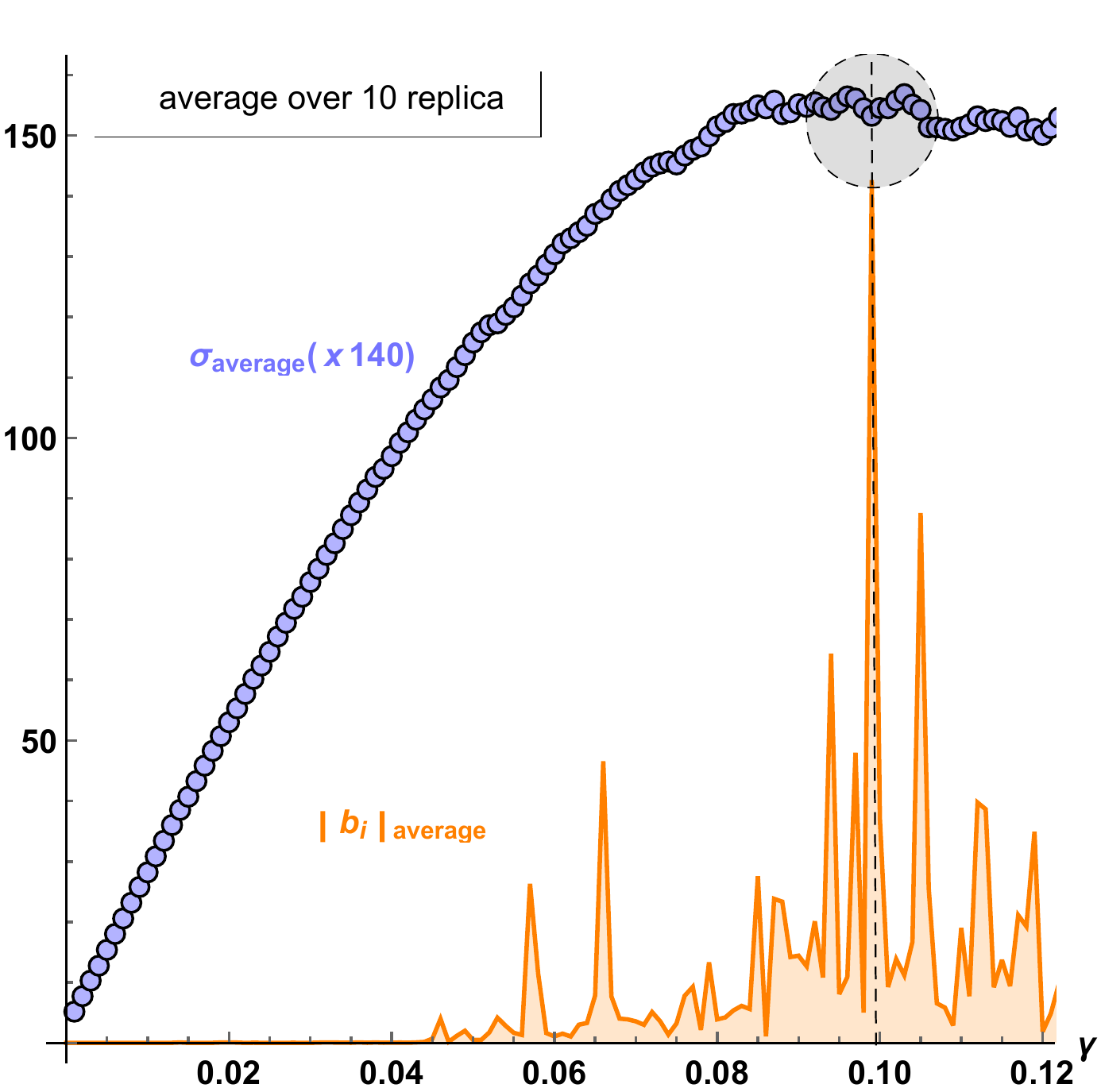}
    \caption{The magnified stress-strain curve (purple circles) and the norm of the Burgers vector $|b_i|$ (orange) averaged over 10 independent replica. The vertical dashed line indicates the location of the main peak. The gray shaded area emphasizes the position of the yielding point.}
    \label{fig:main2}
\end{figure}

In order to make the analysis of the data robust, 10 replicas were created and each was analyzed with stress-strain and Burgers vector analysis of the DTDs. The results are shown in Fig. \ref{fig:main2}. As already anticipated, the norm of the Burgers vector $|b_i|$ averaged over the different replica displays a dominant and sharp peak at the location of the yielding point, around $\gamma \approx 0.1$. As shown explicitly in the Supplemental material \cite{Supplementary}, (I) the norm of the Burgers vector computed on the single replica is able to locate not only the yielding point but also the secondary plastic events manifest in the stress-strain curve as sudden stress fall-off. Strikingly, we observe clear peaks of $|b_i|$ in correspondence of these mechanical instabilities signalled by nearly-zero or slightly negative eigenvalues of the Hessian matrix~\cite{Lerner,Yunjiang}. And, (II) the norm of the Burgers vector is independent of the topology of the closed Burgers loop, This is a manifestation of the topological nature of this object, which ``counts'' the nonaffine displacements inside the close loop, and demonstrates that these DTDs are genuine \emph{topological invariants}.

\begin{figure}[ht]
    \centering
    \includegraphics[width=0.8\linewidth]{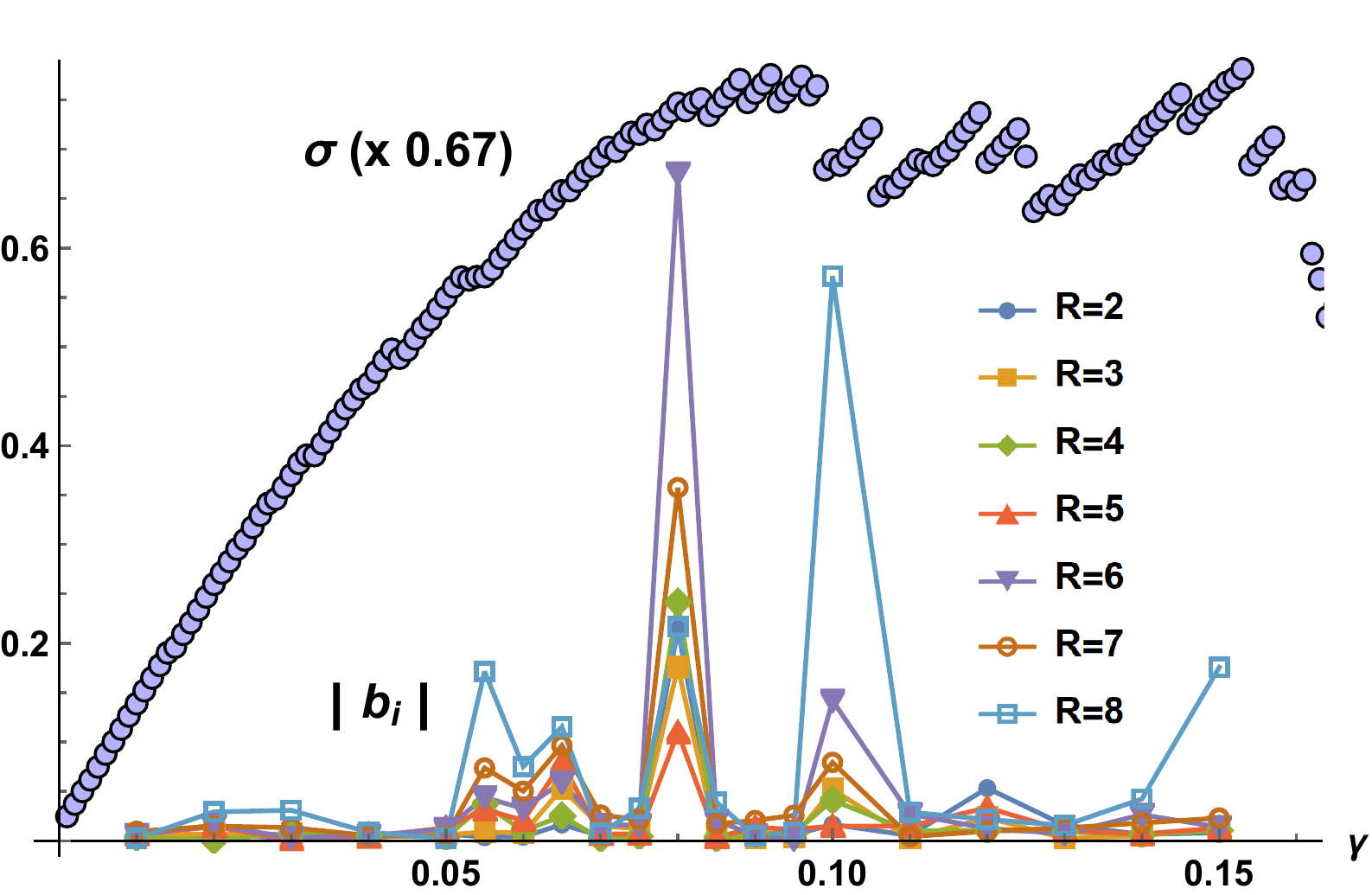}
    
    \vspace{0.2cm}
    
    \includegraphics[width=0.8\linewidth]{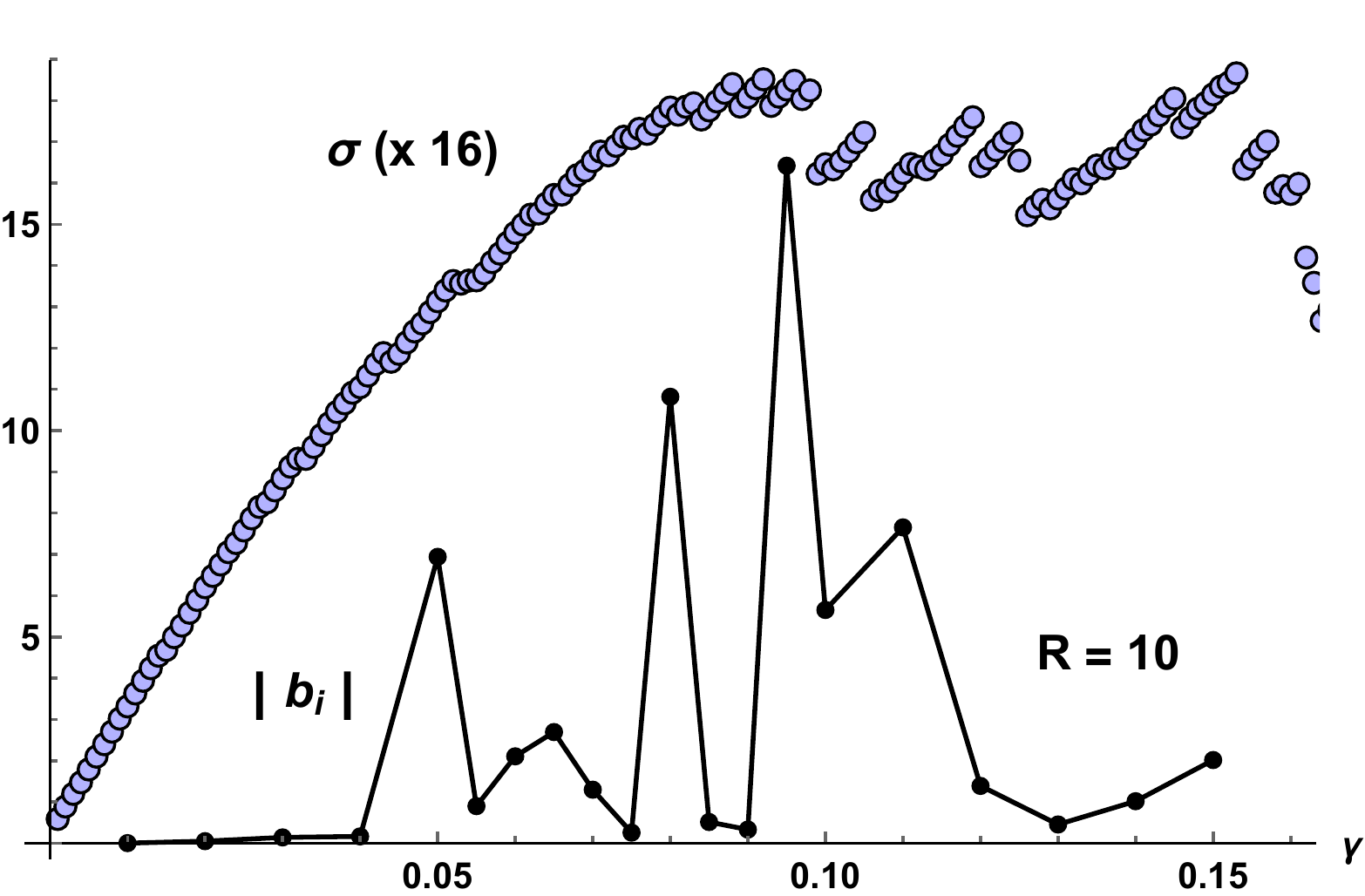}
    \caption{\textbf{Top:} The norm of the Burgers vector $|b_i|$ as a function of the closed loop radius $R$ for a single replica. In purple the corresponding stress-strain curve. \textbf{Bottom: } The same plot with the norm of the Burgers vector for $R=10$.}
    \label{fig:main3}
\end{figure}

\begin{figure}[ht]
    \centering
    \includegraphics[width=1\linewidth]{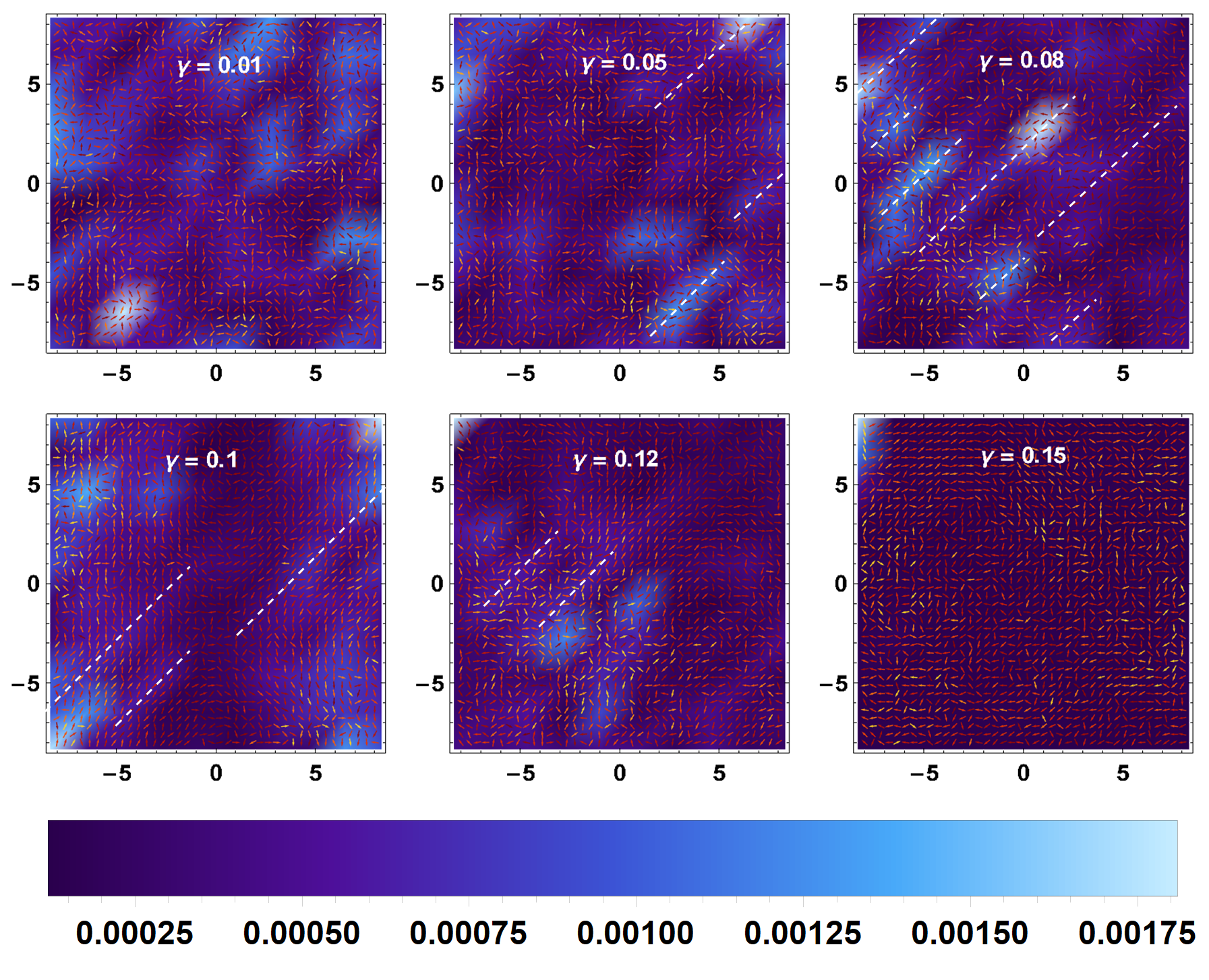}
    \caption{The evolution of the displacements vector $\Vec{u}$ by increasing the external strain $\gamma$. The background color shows the Burgers vector norm and the arrow its direction and local amplitudes. The dashed white lines guide the eye towards the $45^{\circ}$ shear band forming.}
    \label{fig:main4}
\end{figure}

In Fig.~\ref{fig:main3} we present a different analysis of the same data, where now we vary the linear size of the Burgers circuit used to measure the norm $|b_i|$ as a function of strain. This analysis reveals much of the spatial extent of the various plastic events. It is seen that, upon increasing the linear size of the Burgers circuit $\mathcal{L}$ or its radius $R$, the peak of $|b_i|$ corresponding to the yield point $\gamma = 0.1$, grows enormously, much more than the peaks of the plastic events at $\gamma= 0.05-0.06$ and $\gamma= 0.08$, and even more than the post-yielding peaks at $\gamma=0.15$. This fact indicates the formation of a \emph{slip system} spanning the whole material right at yielding, consistent with the formation of shear bands in  Fig.\ref{fig:main4}. A systematic plot of the Burgers peak amplitudes as a function of loop radius $R$ for plastic events at varying $\gamma$ is shown in \cite{Supplementary}.  

Based on the above observations, it is possible to formulate a mechanism of strain-softening and plastic yield in glasses mediated by DTDs and their mutual interaction. After having verified Eq.~\eqref{Burgers} on the basis of the MD simulations, and assuming polar coordinates $(z,\theta)$, the displacement field around a DTDs follows immediately as $u_i=b_i \theta/2\pi$, with the corresponding local elastic strain field being singular, $\epsilon_{\theta z}=\epsilon_{z \theta}= b/4\pi r$~\cite{Kleman}, where $b\equiv |b_i|$ is the modulus of the Burgers vector. 

By simple geometry~\cite{baggioli2021deformations}, one can show that $|b_{i}| \propto |u^{NA}_{i}|$. In turn, from theory, numerical simulations and experiments~\cite{PhysRevE.72.066619,doi:10.1142/S0217984913300020,Laurati,Arratia}, it is known that $|u^{NA}_{i}| \propto \gamma$, where $|u^{NA}_{i}|$ is an average over the sample.
This implies that, due to the nature of nonaffine displacements to grow with $\gamma$, $|b_{i}|$ has, on average, a tendency to grow with the applied strain $\gamma$ as well. This is not exactly what emerges from the single replica shown in the Supplemental information \cite{Supplementary}, where the behaviour of $|b_{i}|$ vs $\gamma$ is rather noisy and intermittent, and occurs mainly through bursts (peaks) in correspondence of major plastic events, and it is these bursts that grow as $\gamma$ increases. 
Although a precise mechanism for DTDs multiplication and growth upon increasing the strain is yet to be identified, it becomes statistically more likely that DTDs begin to interact with each other in the plastic events where $|b_{i}|$ becomes large. In particular, there is an increased likelihood that two DTDs come together with anti-parallel Burgers vectors $\mathbf{b}_{1}$ and $\mathbf{b}_{2}$. It can be shown, using the Peach-K\"{o}hler force, that this gives rise to an attractive interaction force given by~\cite{Landau_elasticity}: 
\begin{equation}
f=-\frac{G \,b_{1}\,b_{2}}{2\pi\, r}\,,\label{force}
\end{equation}
where $b_1$ and $b_2$ are the moduli of the Burgers vectors of the two interacting DTDs and $G$ is the shear modulus. This force is clearly large around the main plastic events where $|b|$ is large.
\\
Hence, it is possible to have a mechanism whereby the rate of encounter and ``coagulation'' between two DTDs with anti-parallel Burgers vector becomes large. DTDs therefore attract each other, with an effective attraction force given by Eq.\eqref{force}, and tend to coagulate into larger aggregates in correspondence of plastic events. This shear-induced aggregation process eventually leads to the formation of slip systems (i.e. shear bands), as the strain increases.

By leveraging these concepts, it is also possible to predict the orientation of the slip systems.
Let $\sigma = F/A_{0}$ be the tensile stress acting on the sample, for example a uniaxial stress, with $F$ the applied tensile force and $A_0$ the sample cross-section area. Denoting with $\phi$ the angle between the normal to the slip plane and the direction of the tensile force $F$, and with $\lambda$ the angle between the slip direction and the direction of $F$, the slip plane area is thus given by $A_{s}=A_{0}/\cos \phi$. Hence the tensile force resolved in the slip direction, $F \cos \lambda$, gives rise to a resolved shear stress given by the well-known Schmid's law~\cite{Schmid1935,Courtney}:
\begin{equation}
\sigma_{RSS}=\sigma \cos \phi \cos \lambda.
\end{equation}
In general, the three directions are not coplanar, hence $\phi + \lambda \neq 90^{o}$, while $\phi + \lambda = 90^{o}$ is the minimum possible value~\cite{Schmid1935,Courtney}. DTDs will, in general, aggregate into slip bands that are oriented randomly. For a given $\sigma$, slip systems will therefore be initiated by facilitated motion of DTDs that self-organize in a slip plane which experiences the largest resolved shear stress $\sigma_{RSS}$, similar to what happens with avalanches that initiate in a spatial direction where the resolved stress is largest and thus can overwhelm frictional resisting forces. The largest resolved stress clearly corresponds to the maximum value of $\cos \phi \cos \lambda$. Under the constraint $\min (\phi +\lambda)=90^{o}$, this happens for $\phi =\lambda = 45^{o}$. Hence, for a uniaxial deformation or for a simple shear deformation, shear bands due to aggregation of DTDs will form at an angle of $45^{o}$ with respect to the tensile axis as observed in our MD simulations, Fig.\ref{fig:main4}, as well as other simulations and experiments~\cite{Procaccia,Wilde,Sopu}. \\
\\
In summary, we have shown that nonaffine displacements in the deformation of amorphous materials lead all the way to the formation of topological singularities (DTDs) in the displacement field, which can be quantitatively characterized by Burgers vectors. We have demonstrated that DTDs are responsible for plastic events on the example of athermal quasistatic shear of model polymer glasses quenched at low temperature. The spatially averaged norm of the Burgers vectors displays peaks corresponding to the plastic events, and an extremely evident non-local (system-spanning) peak at the yield point. Treating DTDs in analogy to dislocations may allow one to formulate a self-consistent mechanism of slip band formation due to the attractive force between anti-parallel DTDs and due to their growing population upon increasing the strain. The preferential alignment of coagulated DTDs (shear bands) along the $45^{\circ}$ degrees direction with respect to the tensile axis is predicted by the Schmid's law, in agreement with all experiments and numerical simulations.
This work provides the quantitative identification of the long-sought ``defects'' which mediate fluidity and plasticity in amorphous solids~\cite{Benzi}. Different from crystals, and from earlier work on glasses~\cite{Steinhardt1979}, the dislocation-like topological defects are not to be found in the static structure but, crucially, in the displacement field under deformation. Furthermore, they originate directly from nonaffine displacements~\cite{PhysRevE.72.066619,Lemaitre,Zaccone2011}. Since the nonaffine displacements in turn originate from the locally low degree of centrosymmetry in the \emph{static} structure of amorphous systems, which is a quantifiable~\cite{Milkus,Plimpton}, this finding opens up the way for identifying the structural signatures of plasticity in glasses~\cite{Schall,Liu,Manning}, but now in terms of atomistically well-defined quantities. Furthermore, it can provide a metric to better distinguish ductile from brittle  first-order like failure~\cite{RossoPNAS,Schall}.

Finally, this work provides a quantitative identification of topological effects in amorphous systems \cite{grushin2020topological} leading to a new geometrical description of plasticity and deformations in glasses. This has potential to open new directions in the chase for ``order'' in disordered systems.

\subsection*{Acknowledgments} 
We thank Michael Landry for discussions and collaboration on related ideas and Giorgio Torrieri, Zohar Nussinov and Yun-Jiang Wang for fruitful discussions and helpful comments. M.B. acknowledges the support of the  Shanghai Municipal Science and Technology Major Project (Grant No.2019SHZDZX01). A.Z. and I.K. acknowledge financial support from US Army Research Laboratory and US Army Research Office through contract nr. W911NF-19-2-0055. 
\bibliographystyle{apsrev4-1}

\bibliography{reale}

\begin{thebibliography}{60}%
\makeatletter
\providecommand \@ifxundefined [1]{%
 \@ifx{#1\undefined}
}%
\providecommand \@ifnum [1]{%
 \ifnum #1\expandafter \@firstoftwo
 \else \expandafter \@secondoftwo
 \fi
}%
\providecommand \@ifx [1]{%
 \ifx #1\expandafter \@firstoftwo
 \else \expandafter \@secondoftwo
 \fi
}%
\providecommand \natexlab [1]{#1}%
\providecommand \enquote  [1]{``#1''}%
\providecommand \bibnamefont  [1]{#1}%
\providecommand \bibfnamefont [1]{#1}%
\providecommand \citenamefont [1]{#1}%
\providecommand \href@noop [0]{\@secondoftwo}%
\providecommand \href [0]{\begingroup \@sanitize@url \@href}%
\providecommand \@href[1]{\@@startlink{#1}\@@href}%
\providecommand \@@href[1]{\endgroup#1\@@endlink}%
\providecommand \@sanitize@url [0]{\catcode `\\12\catcode `\$12\catcode
  `\&12\catcode `\#12\catcode `\^12\catcode `\_12\catcode `\%12\relax}%
\providecommand \@@startlink[1]{}%
\providecommand \@@endlink[0]{}%
\providecommand \url  [0]{\begingroup\@sanitize@url \@url }%
\providecommand \@url [1]{\endgroup\@href {#1}{\urlprefix }}%
\providecommand \urlprefix  [0]{URL }%
\providecommand \Eprint [0]{\href }%
\providecommand \doibase [0]{http://dx.doi.org/}%
\providecommand \selectlanguage [0]{\@gobble}%
\providecommand \bibinfo  [0]{\@secondoftwo}%
\providecommand \bibfield  [0]{\@secondoftwo}%
\providecommand \translation [1]{[#1]}%
\providecommand \BibitemOpen [0]{}%
\providecommand \bibitemStop [0]{}%
\providecommand \bibitemNoStop [0]{.\EOS\space}%
\providecommand \EOS [0]{\spacefactor3000\relax}%
\providecommand \BibitemShut  [1]{\csname bibitem#1\endcsname}%
\let\auto@bib@innerbib\@empty
\bibitem [{\citenamefont {Taylor}(1934)}]{Taylor}%
  \BibitemOpen
  \bibfield  {author} {\bibinfo {author} {\bibfnamefont {G.~I.}\ \bibnamefont
  {Taylor}},\ }\href@noop {} {\bibfield  {journal} {\bibinfo  {journal}
  {Proceedings of the Royal Society of London. Series A, Containing Papers of a
  Mathematical and Physical Character}\ }\textbf {\bibinfo {volume} {145}},\
  \bibinfo {pages} {362} (\bibinfo {year} {1934})}\BibitemShut {NoStop}%
\bibitem [{\citenamefont {Polanyi}(1934)}]{Polanyi}%
  \BibitemOpen
  \bibfield  {author} {\bibinfo {author} {\bibfnamefont {M.}~\bibnamefont
  {Polanyi}},\ }\href {\doibase 10.1007/BF01341481} {\bibfield  {journal}
  {\bibinfo  {journal} {Zeitschrift f{\"u}r Physik}\ }\textbf {\bibinfo
  {volume} {89}},\ \bibinfo {pages} {660} (\bibinfo {year} {1934})}\BibitemShut
  {NoStop}%
\bibitem [{\citenamefont {Orowan}(1934)}]{Orowan}%
  \BibitemOpen
  \bibfield  {author} {\bibinfo {author} {\bibfnamefont {E.}~\bibnamefont
  {Orowan}},\ }\href {\doibase 10.1007/BF01341478} {\bibfield  {journal}
  {\bibinfo  {journal} {Zeitschrift f{\"u}r Physik}\ }\textbf {\bibinfo
  {volume} {89}},\ \bibinfo {pages} {605} (\bibinfo {year} {1934})}\BibitemShut
  {NoStop}%
\bibitem [{\citenamefont {Falk}\ and\ \citenamefont {Langer}(1998)}]{Falk}%
  \BibitemOpen
  \bibfield  {author} {\bibinfo {author} {\bibfnamefont {M.~L.}\ \bibnamefont
  {Falk}}\ and\ \bibinfo {author} {\bibfnamefont {J.~S.}\ \bibnamefont
  {Langer}},\ }\href {\doibase 10.1103/PhysRevE.57.7192} {\bibfield  {journal}
  {\bibinfo  {journal} {Phys. Rev. E}\ }\textbf {\bibinfo {volume} {57}},\
  \bibinfo {pages} {7192} (\bibinfo {year} {1998})}\BibitemShut {NoStop}%
\bibitem [{\citenamefont {Lema{\^i}tre}\ and\ \citenamefont
  {Maloney}(2006{\natexlab{a}})}]{Lemaitre}%
  \BibitemOpen
  \bibfield  {author} {\bibinfo {author} {\bibfnamefont {A.}~\bibnamefont
  {Lema{\^i}tre}}\ and\ \bibinfo {author} {\bibfnamefont {C.}~\bibnamefont
  {Maloney}},\ }\href {\doibase 10.1007/s10955-005-9015-5} {\bibfield
  {journal} {\bibinfo  {journal} {Journal of Statistical Physics}\ }\textbf
  {\bibinfo {volume} {123}},\ \bibinfo {pages} {415} (\bibinfo {year}
  {2006}{\natexlab{a}})}\BibitemShut {NoStop}%
\bibitem [{\citenamefont {Zaccone}\ and\ \citenamefont
  {Scossa-Romano}(2011)}]{Zaccone2011}%
  \BibitemOpen
  \bibfield  {author} {\bibinfo {author} {\bibfnamefont {A.}~\bibnamefont
  {Zaccone}}\ and\ \bibinfo {author} {\bibfnamefont {E.}~\bibnamefont
  {Scossa-Romano}},\ }\href {\doibase 10.1103/PhysRevB.83.184205} {\bibfield
  {journal} {\bibinfo  {journal} {Phys. Rev. B}\ }\textbf {\bibinfo {volume}
  {83}},\ \bibinfo {pages} {184205} (\bibinfo {year} {2011})}\BibitemShut
  {NoStop}%
\bibitem [{\citenamefont {Dasgupta}\ \emph {et~al.}(2012)\citenamefont
  {Dasgupta}, \citenamefont {Hentschel},\ and\ \citenamefont
  {Procaccia}}]{Procaccia}%
  \BibitemOpen
  \bibfield  {author} {\bibinfo {author} {\bibfnamefont {R.}~\bibnamefont
  {Dasgupta}}, \bibinfo {author} {\bibfnamefont {H.~G.~E.}\ \bibnamefont
  {Hentschel}}, \ and\ \bibinfo {author} {\bibfnamefont {I.}~\bibnamefont
  {Procaccia}},\ }\href {\doibase 10.1103/PhysRevLett.109.255502} {\bibfield
  {journal} {\bibinfo  {journal} {Phys. Rev. Lett.}\ }\textbf {\bibinfo
  {volume} {109}},\ \bibinfo {pages} {255502} (\bibinfo {year}
  {2012})}\BibitemShut {NoStop}%
\bibitem [{\citenamefont {De~Giuli}(2020)}]{DeGiuli}%
  \BibitemOpen
  \bibfield  {author} {\bibinfo {author} {\bibfnamefont {E.}~\bibnamefont
  {De~Giuli}},\ }\href {\doibase 10.1103/PhysRevE.101.043002} {\bibfield
  {journal} {\bibinfo  {journal} {Phys. Rev. E}\ }\textbf {\bibinfo {volume}
  {101}},\ \bibinfo {pages} {043002} (\bibinfo {year} {2020})}\BibitemShut
  {NoStop}%
\bibitem [{\citenamefont {Hieronymus-Schmidt}\ \emph
  {et~al.}(2017)\citenamefont {Hieronymus-Schmidt}, \citenamefont {R\"osner},
  \citenamefont {Wilde},\ and\ \citenamefont {Zaccone}}]{Wilde}%
  \BibitemOpen
  \bibfield  {author} {\bibinfo {author} {\bibfnamefont {V.}~\bibnamefont
  {Hieronymus-Schmidt}}, \bibinfo {author} {\bibfnamefont {H.}~\bibnamefont
  {R\"osner}}, \bibinfo {author} {\bibfnamefont {G.}~\bibnamefont {Wilde}}, \
  and\ \bibinfo {author} {\bibfnamefont {A.}~\bibnamefont {Zaccone}},\ }\href
  {\doibase 10.1103/PhysRevB.95.134111} {\bibfield  {journal} {\bibinfo
  {journal} {Phys. Rev. B}\ }\textbf {\bibinfo {volume} {95}},\ \bibinfo
  {pages} {134111} (\bibinfo {year} {2017})}\BibitemShut {NoStop}%
\bibitem [{\citenamefont {\ifmmode~\mbox{\c{S}}\else \c{S}\fi{}opu}\ \emph
  {et~al.}(2017)\citenamefont {\ifmmode~\mbox{\c{S}}\else \c{S}\fi{}opu},
  \citenamefont {Stukowski}, \citenamefont {Stoica},\ and\ \citenamefont
  {Scudino}}]{Sopu}%
  \BibitemOpen
  \bibfield  {author} {\bibinfo {author} {\bibfnamefont {D.}~\bibnamefont
  {\ifmmode~\mbox{\c{S}}\else \c{S}\fi{}opu}}, \bibinfo {author} {\bibfnamefont
  {A.}~\bibnamefont {Stukowski}}, \bibinfo {author} {\bibfnamefont
  {M.}~\bibnamefont {Stoica}}, \ and\ \bibinfo {author} {\bibfnamefont
  {S.}~\bibnamefont {Scudino}},\ }\href {\doibase
  10.1103/PhysRevLett.119.195503} {\bibfield  {journal} {\bibinfo  {journal}
  {Phys. Rev. Lett.}\ }\textbf {\bibinfo {volume} {119}},\ \bibinfo {pages}
  {195503} (\bibinfo {year} {2017})}\BibitemShut {NoStop}%
\bibitem [{\citenamefont {Cao}\ \emph {et~al.}(2018)\citenamefont {Cao},
  \citenamefont {Nicolas}, \citenamefont {Trimcev},\ and\ \citenamefont
  {Rosso}}]{Rosso}%
  \BibitemOpen
  \bibfield  {author} {\bibinfo {author} {\bibfnamefont {X.}~\bibnamefont
  {Cao}}, \bibinfo {author} {\bibfnamefont {A.}~\bibnamefont {Nicolas}},
  \bibinfo {author} {\bibfnamefont {D.}~\bibnamefont {Trimcev}}, \ and\
  \bibinfo {author} {\bibfnamefont {A.}~\bibnamefont {Rosso}},\ }\href
  {\doibase 10.1039/C7SM02510F} {\bibfield  {journal} {\bibinfo  {journal}
  {Soft Matter}\ }\textbf {\bibinfo {volume} {14}},\ \bibinfo {pages} {3640}
  (\bibinfo {year} {2018})}\BibitemShut {NoStop}%
\bibitem [{\citenamefont {Ghosh}\ \emph {et~al.}(2017)\citenamefont {Ghosh},
  \citenamefont {Budrikis}, \citenamefont {Chikkadi}, \citenamefont {Sellerio},
  \citenamefont {Zapperi},\ and\ \citenamefont {Schall}}]{Zapperi}%
  \BibitemOpen
  \bibfield  {author} {\bibinfo {author} {\bibfnamefont {A.}~\bibnamefont
  {Ghosh}}, \bibinfo {author} {\bibfnamefont {Z.}~\bibnamefont {Budrikis}},
  \bibinfo {author} {\bibfnamefont {V.}~\bibnamefont {Chikkadi}}, \bibinfo
  {author} {\bibfnamefont {A.~L.}\ \bibnamefont {Sellerio}}, \bibinfo {author}
  {\bibfnamefont {S.}~\bibnamefont {Zapperi}}, \ and\ \bibinfo {author}
  {\bibfnamefont {P.}~\bibnamefont {Schall}},\ }\href {\doibase
  10.1103/PhysRevLett.118.148001} {\bibfield  {journal} {\bibinfo  {journal}
  {Phys. Rev. Lett.}\ }\textbf {\bibinfo {volume} {118}},\ \bibinfo {pages}
  {148001} (\bibinfo {year} {2017})}\BibitemShut {NoStop}%
\bibitem [{\citenamefont {Chaudhari}\ \emph {et~al.}(1979)\citenamefont
  {Chaudhari}, \citenamefont {Levi},\ and\ \citenamefont
  {Steinhardt}}]{Steinhardt1979}%
  \BibitemOpen
  \bibfield  {author} {\bibinfo {author} {\bibfnamefont {P.}~\bibnamefont
  {Chaudhari}}, \bibinfo {author} {\bibfnamefont {A.}~\bibnamefont {Levi}}, \
  and\ \bibinfo {author} {\bibfnamefont {P.}~\bibnamefont {Steinhardt}},\
  }\href {\doibase 10.1103/PhysRevLett.43.1517} {\bibfield  {journal} {\bibinfo
   {journal} {Phys. Rev. Lett.}\ }\textbf {\bibinfo {volume} {43}},\ \bibinfo
  {pages} {1517} (\bibinfo {year} {1979})}\BibitemShut {NoStop}%
\bibitem [{\citenamefont {Steinhardt}\ and\ \citenamefont
  {Chaudhari}(1981)}]{Steinhardt1981}%
  \BibitemOpen
  \bibfield  {author} {\bibinfo {author} {\bibfnamefont {P.~J.}\ \bibnamefont
  {Steinhardt}}\ and\ \bibinfo {author} {\bibfnamefont {P.}~\bibnamefont
  {Chaudhari}},\ }\href {\doibase 10.1080/01418618108235816} {\bibfield
  {journal} {\bibinfo  {journal} {Philosophical Magazine A}\ }\textbf {\bibinfo
  {volume} {44}},\ \bibinfo {pages} {1375} (\bibinfo {year} {1981})},\ \Eprint
  {http://arxiv.org/abs/https://doi.org/10.1080/01418618108235816}
  {https://doi.org/10.1080/01418618108235816} \BibitemShut {NoStop}%
\bibitem [{\citenamefont {Acharya}\ and\ \citenamefont {Widom}(2017)}]{Widom}%
  \BibitemOpen
  \bibfield  {author} {\bibinfo {author} {\bibfnamefont {A.}~\bibnamefont
  {Acharya}}\ and\ \bibinfo {author} {\bibfnamefont {M.}~\bibnamefont
  {Widom}},\ }\href {\doibase https://doi.org/10.1016/j.jmps.2017.03.014}
  {\bibfield  {journal} {\bibinfo  {journal} {Journal of the Mechanics and
  Physics of Solids}\ }\textbf {\bibinfo {volume} {104}},\ \bibinfo {pages} {1
  } (\bibinfo {year} {2017})}\BibitemShut {NoStop}%
\bibitem [{\citenamefont {Moshe}\ \emph {et~al.}(2015)\citenamefont {Moshe},
  \citenamefont {Levin}, \citenamefont {Aharoni}, \citenamefont {Kupferman},\
  and\ \citenamefont {Sharon}}]{Moshe}%
  \BibitemOpen
  \bibfield  {author} {\bibinfo {author} {\bibfnamefont {M.}~\bibnamefont
  {Moshe}}, \bibinfo {author} {\bibfnamefont {I.}~\bibnamefont {Levin}},
  \bibinfo {author} {\bibfnamefont {H.}~\bibnamefont {Aharoni}}, \bibinfo
  {author} {\bibfnamefont {R.}~\bibnamefont {Kupferman}}, \ and\ \bibinfo
  {author} {\bibfnamefont {E.}~\bibnamefont {Sharon}},\ }\href {\doibase
  10.1073/pnas.1506531112} {\bibfield  {journal} {\bibinfo  {journal}
  {Proceedings of the National Academy of Sciences}\ }\textbf {\bibinfo
  {volume} {112}},\ \bibinfo {pages} {10873} (\bibinfo {year} {2015})},\
  \Eprint
  {http://arxiv.org/abs/https://www.pnas.org/content/112/35/10873.full.pdf}
  {https://www.pnas.org/content/112/35/10873.full.pdf} \BibitemShut {NoStop}%
\bibitem [{\citenamefont {Chaikin}\ and\ \citenamefont
  {Lubensky}(2000)}]{chaikin2000principles}%
  \BibitemOpen
  \bibfield  {author} {\bibinfo {author} {\bibfnamefont {P.}~\bibnamefont
  {Chaikin}}\ and\ \bibinfo {author} {\bibfnamefont {T.}~\bibnamefont
  {Lubensky}},\ }\href {https://books.google.gr/books?id=P9YjNjzr9OIC} {\emph
  {\bibinfo {title} {Principles of Condensed Matter Physics}}}\ (\bibinfo
  {publisher} {Cambridge University Press},\ \bibinfo {year}
  {2000})\BibitemShut {NoStop}%
\bibitem [{\citenamefont {Landau}\ and\ \citenamefont
  {Lifshitz}(1986)}]{Landau_elasticity}%
  \BibitemOpen
  \bibfield  {author} {\bibinfo {author} {\bibfnamefont {L.}~\bibnamefont
  {Landau}}\ and\ \bibinfo {author} {\bibfnamefont {E.}~\bibnamefont
  {Lifshitz}},\ }\href@noop {} {\emph {\bibinfo {title} {Theory of Elasticity:
  Volume 6}}}\ (\bibinfo  {publisher} {Pergamon Press},\ \bibinfo {year}
  {1986})\BibitemShut {NoStop}%
\bibitem [{\citenamefont {DiDonna}\ and\ \citenamefont
  {Lubensky}(2005)}]{PhysRevE.72.066619}%
  \BibitemOpen
  \bibfield  {author} {\bibinfo {author} {\bibfnamefont {B.~A.}\ \bibnamefont
  {DiDonna}}\ and\ \bibinfo {author} {\bibfnamefont {T.~C.}\ \bibnamefont
  {Lubensky}},\ }\href {\doibase 10.1103/PhysRevE.72.066619} {\bibfield
  {journal} {\bibinfo  {journal} {Phys. Rev. E}\ }\textbf {\bibinfo {volume}
  {72}},\ \bibinfo {pages} {066619} (\bibinfo {year} {2005})}\BibitemShut
  {NoStop}%
\bibitem [{\citenamefont {Cui}\ \emph {et~al.}(2019)\citenamefont {Cui},
  \citenamefont {Zaccone},\ and\ \citenamefont {Rodney}}]{Cui}%
  \BibitemOpen
  \bibfield  {author} {\bibinfo {author} {\bibfnamefont {B.}~\bibnamefont
  {Cui}}, \bibinfo {author} {\bibfnamefont {A.}~\bibnamefont {Zaccone}}, \ and\
  \bibinfo {author} {\bibfnamefont {D.}~\bibnamefont {Rodney}},\ }\href
  {\doibase 10.1063/1.5129025} {\bibfield  {journal} {\bibinfo  {journal} {The
  Journal of Chemical Physics}\ }\textbf {\bibinfo {volume} {151}},\ \bibinfo
  {pages} {224509} (\bibinfo {year} {2019})},\ \Eprint
  {http://arxiv.org/abs/https://doi.org/10.1063/1.5129025}
  {https://doi.org/10.1063/1.5129025} \BibitemShut {NoStop}%
\bibitem [{\citenamefont {Love}(1892)}]{M.1906}%
  \BibitemOpen
  \bibfield  {author} {\bibinfo {author} {\bibfnamefont {A.~E.~H.}\
  \bibnamefont {Love}},\ }\href@noop {} {\emph {\bibinfo {title} {A Treatise on
  the Mathematical Theory of Elasticity}}}\ (\bibinfo  {publisher} {Cambridge
  University Press, Cambridge},\ \bibinfo {year} {1892})\BibitemShut {NoStop}%
\bibitem [{\citenamefont {Beltrami}(1886)}]{beltrami1886sull}%
  \BibitemOpen
  \bibfield  {author} {\bibinfo {author} {\bibfnamefont {E.}~\bibnamefont
  {Beltrami}},\ }\href@noop {} {\emph {\bibinfo {title} {Sull'interpretazione
  meccanica delle formole di Maxwell: memoria}}}\ (\bibinfo  {publisher}
  {Tipografia Gamberini e Parmeggiani},\ \bibinfo {year} {1886})\BibitemShut
  {NoStop}%
\bibitem [{\citenamefont {Kleinert}(1989)}]{kleinert1989gauge}%
  \BibitemOpen
  \bibfield  {author} {\bibinfo {author} {\bibfnamefont {H.}~\bibnamefont
  {Kleinert}},\ }\href {https://books.google.es/books?id=jshUAAAAYAAJ} {\emph
  {\bibinfo {title} {Gauge Fields in Condensed Matter}}},\ \bibinfo {series}
  {Gauge Fields in Condensed Matter}\ No.\ \bibinfo {number} {v. 2}\ (\bibinfo
  {publisher} {World Scientific},\ \bibinfo {year} {1989})\BibitemShut
  {NoStop}%
\bibitem [{\citenamefont {Grozdanov}\ and\ \citenamefont
  {Poovuttikul}(2019)}]{Grozdanov:2017kyl}%
  \BibitemOpen
  \bibfield  {author} {\bibinfo {author} {\bibfnamefont {S.}~\bibnamefont
  {Grozdanov}}\ and\ \bibinfo {author} {\bibfnamefont {N.}~\bibnamefont
  {Poovuttikul}},\ }\href {\doibase 10.1007/JHEP04(2019)141} {\bibfield
  {journal} {\bibinfo  {journal} {JHEP}\ }\textbf {\bibinfo {volume} {04}},\
  \bibinfo {pages} {141} (\bibinfo {year} {2019})},\ \Eprint
  {http://arxiv.org/abs/1707.04182} {arXiv:1707.04182 [hep-th]} \BibitemShut
  {NoStop}%
\bibitem [{\citenamefont {Baggioli}\ \emph {et~al.}(2021)\citenamefont
  {Baggioli}, \citenamefont {Landry},\ and\ \citenamefont
  {Zaccone}}]{baggioli2021deformations}%
  \BibitemOpen
  \bibfield  {author} {\bibinfo {author} {\bibfnamefont {M.}~\bibnamefont
  {Baggioli}}, \bibinfo {author} {\bibfnamefont {M.}~\bibnamefont {Landry}}, \
  and\ \bibinfo {author} {\bibfnamefont {A.}~\bibnamefont {Zaccone}},\
  }\href@noop {} {\enquote {\bibinfo {title} {Deformations, relaxation and
  broken symmetries in liquids, solids and glasses: a unified topological field
  theory},}\ } (\bibinfo {year} {2021}),\ \Eprint
  {http://arxiv.org/abs/2101.05015} {arXiv:2101.05015 [cond-mat.soft]}
  \BibitemShut {NoStop}%
\bibitem [{\citenamefont {Zaanen}\ \emph {et~al.}(2004)\citenamefont {Zaanen},
  \citenamefont {Nussinov},\ and\ \citenamefont {Mukhin}}]{Nussinov1}%
  \BibitemOpen
  \bibfield  {author} {\bibinfo {author} {\bibfnamefont {J.}~\bibnamefont
  {Zaanen}}, \bibinfo {author} {\bibfnamefont {Z.}~\bibnamefont {Nussinov}}, \
  and\ \bibinfo {author} {\bibfnamefont {S.}~\bibnamefont {Mukhin}},\ }\href
  {\doibase https://doi.org/10.1016/j.aop.2003.10.003} {\bibfield  {journal}
  {\bibinfo  {journal} {Annals of Physics}\ }\textbf {\bibinfo {volume}
  {310}},\ \bibinfo {pages} {181} (\bibinfo {year} {2004})}\BibitemShut
  {NoStop}%
\bibitem [{\citenamefont {Grozdanov}\ \emph {et~al.}(2017)\citenamefont
  {Grozdanov}, \citenamefont {Hofman},\ and\ \citenamefont
  {Iqbal}}]{Grozdanov:2016tdf}%
  \BibitemOpen
  \bibfield  {author} {\bibinfo {author} {\bibfnamefont {S.}~\bibnamefont
  {Grozdanov}}, \bibinfo {author} {\bibfnamefont {D.~M.}\ \bibnamefont
  {Hofman}}, \ and\ \bibinfo {author} {\bibfnamefont {N.}~\bibnamefont
  {Iqbal}},\ }\href {\doibase 10.1103/PhysRevD.95.096003} {\bibfield  {journal}
  {\bibinfo  {journal} {Phys. Rev. D}\ }\textbf {\bibinfo {volume} {95}},\
  \bibinfo {pages} {096003} (\bibinfo {year} {2017})},\ \Eprint
  {http://arxiv.org/abs/1610.07392} {arXiv:1610.07392 [hep-th]} \BibitemShut
  {NoStop}%
\bibitem [{\citenamefont {Grozdanov}\ and\ \citenamefont
  {Poovuttikul}(2018)}]{Grozdanov:2018ewh}%
  \BibitemOpen
  \bibfield  {author} {\bibinfo {author} {\bibfnamefont {S.}~\bibnamefont
  {Grozdanov}}\ and\ \bibinfo {author} {\bibfnamefont {N.}~\bibnamefont
  {Poovuttikul}},\ }\href {\doibase 10.1103/PhysRevD.97.106005} {\bibfield
  {journal} {\bibinfo  {journal} {Phys. Rev. D}\ }\textbf {\bibinfo {volume}
  {97}},\ \bibinfo {pages} {106005} (\bibinfo {year} {2018})},\ \Eprint
  {http://arxiv.org/abs/1801.03199} {arXiv:1801.03199 [hep-th]} \BibitemShut
  {NoStop}%
\bibitem [{\citenamefont {Beekman}\ \emph {et~al.}(2017)\citenamefont
  {Beekman}, \citenamefont {Nissinen}, \citenamefont {Wu}, \citenamefont {Liu},
  \citenamefont {Slager}, \citenamefont {Nussinov}, \citenamefont {Cvetkovic},\
  and\ \citenamefont {Zaanen}}]{BEEKMAN20171}%
  \BibitemOpen
  \bibfield  {author} {\bibinfo {author} {\bibfnamefont {A.~J.}\ \bibnamefont
  {Beekman}}, \bibinfo {author} {\bibfnamefont {J.}~\bibnamefont {Nissinen}},
  \bibinfo {author} {\bibfnamefont {K.}~\bibnamefont {Wu}}, \bibinfo {author}
  {\bibfnamefont {K.}~\bibnamefont {Liu}}, \bibinfo {author} {\bibfnamefont
  {R.-J.}\ \bibnamefont {Slager}}, \bibinfo {author} {\bibfnamefont
  {Z.}~\bibnamefont {Nussinov}}, \bibinfo {author} {\bibfnamefont
  {V.}~\bibnamefont {Cvetkovic}}, \ and\ \bibinfo {author} {\bibfnamefont
  {J.}~\bibnamefont {Zaanen}},\ }\href {\doibase
  https://doi.org/10.1016/j.physrep.2017.03.004} {\bibfield  {journal}
  {\bibinfo  {journal} {Physics Reports}\ }\textbf {\bibinfo {volume} {683}},\
  \bibinfo {pages} {1} (\bibinfo {year} {2017})},\ \bibinfo {note} {dual gauge
  field theory of quantum liquid crystals in two dimensions}\BibitemShut
  {NoStop}%
\bibitem [{\citenamefont {Cvetkovic}\ \emph {et~al.}(2006)\citenamefont
  {Cvetkovic}, \citenamefont {Nussinov},\ and\ \citenamefont
  {Zaanen}}]{doi:10.1080/14786430600636328}%
  \BibitemOpen
  \bibfield  {author} {\bibinfo {author} {\bibfnamefont {V.}~\bibnamefont
  {Cvetkovic}}, \bibinfo {author} {\bibfnamefont {Z.}~\bibnamefont {Nussinov}},
  \ and\ \bibinfo {author} {\bibfnamefont {J.}~\bibnamefont {Zaanen}},\ }\href
  {\doibase 10.1080/14786430600636328} {\bibfield  {journal} {\bibinfo
  {journal} {Philosophical Magazine}\ }\textbf {\bibinfo {volume} {86}},\
  \bibinfo {pages} {2995} (\bibinfo {year} {2006})},\ \Eprint
  {http://arxiv.org/abs/https://doi.org/10.1080/14786430600636328}
  {https://doi.org/10.1080/14786430600636328} \BibitemShut {NoStop}%
\bibitem [{\citenamefont {{Cvetkovic}}\ \emph {et~al.}(2009)\citenamefont
  {{Cvetkovic}}, \citenamefont {{Nussinov}},\ and\ \citenamefont
  {{Zaanen}}}]{2009arXiv0905.2996C}%
  \BibitemOpen
  \bibfield  {author} {\bibinfo {author} {\bibfnamefont {V.}~\bibnamefont
  {{Cvetkovic}}}, \bibinfo {author} {\bibfnamefont {Z.}~\bibnamefont
  {{Nussinov}}}, \ and\ \bibinfo {author} {\bibfnamefont {J.}~\bibnamefont
  {{Zaanen}}},\ }\href@noop {} {\bibfield  {journal} {\bibinfo  {journal}
  {arXiv e-prints}\ ,\ \bibinfo {eid} {arXiv:0905.2996}} (\bibinfo {year}
  {2009})},\ \Eprint {http://arxiv.org/abs/0905.2996} {arXiv:0905.2996
  [cond-mat.str-el]} \BibitemShut {NoStop}%
\bibitem [{\citenamefont {Ronhovde}\ \emph {et~al.}(2011)\citenamefont
  {Ronhovde}, \citenamefont {Chakrabarty}, \citenamefont {Hu}, \citenamefont
  {Sahu}, \citenamefont {Sahu}, \citenamefont {Kelton}, \citenamefont {Mauro},\
  and\ \citenamefont {Nussinov}}]{Ronhovde2011}%
  \BibitemOpen
  \bibfield  {author} {\bibinfo {author} {\bibfnamefont {P.}~\bibnamefont
  {Ronhovde}}, \bibinfo {author} {\bibfnamefont {S.}~\bibnamefont
  {Chakrabarty}}, \bibinfo {author} {\bibfnamefont {D.}~\bibnamefont {Hu}},
  \bibinfo {author} {\bibfnamefont {M.}~\bibnamefont {Sahu}}, \bibinfo {author}
  {\bibfnamefont {K.~K.}\ \bibnamefont {Sahu}}, \bibinfo {author}
  {\bibfnamefont {K.~F.}\ \bibnamefont {Kelton}}, \bibinfo {author}
  {\bibfnamefont {N.~A.}\ \bibnamefont {Mauro}}, \ and\ \bibinfo {author}
  {\bibfnamefont {Z.}~\bibnamefont {Nussinov}},\ }\href {\doibase
  10.1140/epje/i2011-11105-9} {\bibfield  {journal} {\bibinfo  {journal} {The
  European Physical Journal E}\ }\textbf {\bibinfo {volume} {34}},\ \bibinfo
  {pages} {105} (\bibinfo {year} {2011})}\BibitemShut {NoStop}%
\bibitem [{\citenamefont {Kapustin}\ and\ \citenamefont
  {Thorngren}(2013)}]{Kapustin:2013uxa}%
  \BibitemOpen
  \bibfield  {author} {\bibinfo {author} {\bibfnamefont {A.}~\bibnamefont
  {Kapustin}}\ and\ \bibinfo {author} {\bibfnamefont {R.}~\bibnamefont
  {Thorngren}},\ }\href@noop {} {\  (\bibinfo {year} {2013})},\ \Eprint
  {http://arxiv.org/abs/1309.4721} {arXiv:1309.4721 [hep-th]} \BibitemShut
  {NoStop}%
\bibitem [{\citenamefont {Gaiotto}\ \emph {et~al.}(2015)\citenamefont
  {Gaiotto}, \citenamefont {Kapustin}, \citenamefont {Seiberg},\ and\
  \citenamefont {Willett}}]{Gaiotto:2014kfa}%
  \BibitemOpen
  \bibfield  {author} {\bibinfo {author} {\bibfnamefont {D.}~\bibnamefont
  {Gaiotto}}, \bibinfo {author} {\bibfnamefont {A.}~\bibnamefont {Kapustin}},
  \bibinfo {author} {\bibfnamefont {N.}~\bibnamefont {Seiberg}}, \ and\
  \bibinfo {author} {\bibfnamefont {B.}~\bibnamefont {Willett}},\ }\href
  {\doibase 10.1007/JHEP02(2015)172} {\bibfield  {journal} {\bibinfo  {journal}
  {JHEP}\ }\textbf {\bibinfo {volume} {02}},\ \bibinfo {pages} {172} (\bibinfo
  {year} {2015})},\ \Eprint {http://arxiv.org/abs/1412.5148} {arXiv:1412.5148
  [hep-th]} \BibitemShut {NoStop}%
\bibitem [{\citenamefont {Yoshida}(2016)}]{PhysRevB.93.155131}%
  \BibitemOpen
  \bibfield  {author} {\bibinfo {author} {\bibfnamefont {B.}~\bibnamefont
  {Yoshida}},\ }\href {\doibase 10.1103/PhysRevB.93.155131} {\bibfield
  {journal} {\bibinfo  {journal} {Phys. Rev. B}\ }\textbf {\bibinfo {volume}
  {93}},\ \bibinfo {pages} {155131} (\bibinfo {year} {2016})}\BibitemShut
  {NoStop}%
\bibitem [{\citenamefont {Palyulin}\ \emph {et~al.}(2018)\citenamefont
  {Palyulin}, \citenamefont {Ness}, \citenamefont {Milkus}, \citenamefont
  {Elder}, \citenamefont {Sirk},\ and\ \citenamefont {Zaccone}}]{Palyulin}%
  \BibitemOpen
  \bibfield  {author} {\bibinfo {author} {\bibfnamefont {V.~V.}\ \bibnamefont
  {Palyulin}}, \bibinfo {author} {\bibfnamefont {C.}~\bibnamefont {Ness}},
  \bibinfo {author} {\bibfnamefont {R.}~\bibnamefont {Milkus}}, \bibinfo
  {author} {\bibfnamefont {R.~M.}\ \bibnamefont {Elder}}, \bibinfo {author}
  {\bibfnamefont {T.~W.}\ \bibnamefont {Sirk}}, \ and\ \bibinfo {author}
  {\bibfnamefont {A.}~\bibnamefont {Zaccone}},\ }\href {\doibase
  10.1039/C8SM01468J} {\bibfield  {journal} {\bibinfo  {journal} {Soft Matter}\
  }\textbf {\bibinfo {volume} {14}},\ \bibinfo {pages} {8475} (\bibinfo {year}
  {2018})}\BibitemShut {NoStop}%
\bibitem [{Sup()}]{Supplementary}%
  \BibitemOpen
  \href@noop {} {\enquote {\bibinfo {title} {Supplementary material available
  at...}}\ }\BibitemShut {NoStop}%
\bibitem [{\citenamefont {Kremer}\ and\ \citenamefont
  {Grest}(1986)}]{Kremer1986}%
  \BibitemOpen
  \bibfield  {author} {\bibinfo {author} {\bibfnamefont {K.}~\bibnamefont
  {Kremer}}\ and\ \bibinfo {author} {\bibfnamefont {G.~S.}\ \bibnamefont
  {Grest}},\ }\href@noop {} {\bibfield  {journal} {\bibinfo  {journal} {Phys.
  Rev. A}\ }\textbf {\bibinfo {volume} {33}},\ \bibinfo {pages} {3628}
  (\bibinfo {year} {1986})}\BibitemShut {NoStop}%
\bibitem [{\citenamefont {Kriuchevskyi}\ \emph {et~al.}(2020)\citenamefont
  {Kriuchevskyi}, \citenamefont {Palyulin}, \citenamefont {Milkus},
  \citenamefont {Elder}, \citenamefont {Sirk},\ and\ \citenamefont
  {Zaccone}}]{PhysRevB.102.024108}%
  \BibitemOpen
  \bibfield  {author} {\bibinfo {author} {\bibfnamefont {I.}~\bibnamefont
  {Kriuchevskyi}}, \bibinfo {author} {\bibfnamefont {V.~V.}\ \bibnamefont
  {Palyulin}}, \bibinfo {author} {\bibfnamefont {R.}~\bibnamefont {Milkus}},
  \bibinfo {author} {\bibfnamefont {R.~M.}\ \bibnamefont {Elder}}, \bibinfo
  {author} {\bibfnamefont {T.~W.}\ \bibnamefont {Sirk}}, \ and\ \bibinfo
  {author} {\bibfnamefont {A.}~\bibnamefont {Zaccone}},\ }\href {\doibase
  10.1103/PhysRevB.102.024108} {\bibfield  {journal} {\bibinfo  {journal}
  {Phys. Rev. B}\ }\textbf {\bibinfo {volume} {102}},\ \bibinfo {pages}
  {024108} (\bibinfo {year} {2020})}\BibitemShut {NoStop}%
\bibitem [{\citenamefont {Plimpton}(1995)}]{LAMMPS}%
  \BibitemOpen
  \bibfield  {author} {\bibinfo {author} {\bibfnamefont {S.}~\bibnamefont
  {Plimpton}},\ }\href@noop {} {\bibfield  {journal} {\bibinfo  {journal} {J.
  Comp. Phys}\ }\textbf {\bibinfo {volume} {117}},\ \bibinfo {pages} {1}
  (\bibinfo {year} {1995})},\ \bibinfo {note} {see also:
  http://lammps.sandia.gov}\BibitemShut {NoStop}%
\bibitem [{\citenamefont {Acharya}\ and\ \citenamefont
  {Bassani}(2000)}]{Bassani}%
  \BibitemOpen
  \bibfield  {author} {\bibinfo {author} {\bibfnamefont {A.}~\bibnamefont
  {Acharya}}\ and\ \bibinfo {author} {\bibfnamefont {J.}~\bibnamefont
  {Bassani}},\ }\href {\doibase https://doi.org/10.1016/S0022-5096(99)00075-7}
  {\bibfield  {journal} {\bibinfo  {journal} {Journal of the Mechanics and
  Physics of Solids}\ }\textbf {\bibinfo {volume} {48}},\ \bibinfo {pages}
  {1565 } (\bibinfo {year} {2000})}\BibitemShut {NoStop}%
\bibitem [{\citenamefont {Zimmerman}\ \emph {et~al.}(2009)\citenamefont
  {Zimmerman}, \citenamefont {Bammann},\ and\ \citenamefont {Gao}}]{Zimmerman}%
  \BibitemOpen
  \bibfield  {author} {\bibinfo {author} {\bibfnamefont {J.~A.}\ \bibnamefont
  {Zimmerman}}, \bibinfo {author} {\bibfnamefont {D.~J.}\ \bibnamefont
  {Bammann}}, \ and\ \bibinfo {author} {\bibfnamefont {H.}~\bibnamefont
  {Gao}},\ }\href {\doibase https://doi.org/10.1016/j.ijsolstr.2008.08.036}
  {\bibfield  {journal} {\bibinfo  {journal} {International Journal of Solids
  and Structures}\ }\textbf {\bibinfo {volume} {46}},\ \bibinfo {pages} {238 }
  (\bibinfo {year} {2009})}\BibitemShut {NoStop}%
\bibitem [{\citenamefont {Ruggiero}\ and\ \citenamefont
  {Tartaglia}(2003)}]{Tartaglia}%
  \BibitemOpen
  \bibfield  {author} {\bibinfo {author} {\bibfnamefont {M.~L.}\ \bibnamefont
  {Ruggiero}}\ and\ \bibinfo {author} {\bibfnamefont {A.}~\bibnamefont
  {Tartaglia}},\ }\href {\doibase 10.1119/1.1596176} {\bibfield  {journal}
  {\bibinfo  {journal} {American Journal of Physics}\ }\textbf {\bibinfo
  {volume} {71}},\ \bibinfo {pages} {1303} (\bibinfo {year} {2003})},\ \Eprint
  {http://arxiv.org/abs/https://doi.org/10.1119/1.1596176}
  {https://doi.org/10.1119/1.1596176} \BibitemShut {NoStop}%
\bibitem [{\citenamefont {Gartner}\ and\ \citenamefont
  {Lerner}(2016)}]{Lerner}%
  \BibitemOpen
  \bibfield  {author} {\bibinfo {author} {\bibfnamefont {L.}~\bibnamefont
  {Gartner}}\ and\ \bibinfo {author} {\bibfnamefont {E.}~\bibnamefont
  {Lerner}},\ }\href {\doibase 10.1103/PhysRevE.93.011001} {\bibfield
  {journal} {\bibinfo  {journal} {Phys. Rev. E}\ }\textbf {\bibinfo {volume}
  {93}},\ \bibinfo {pages} {011001} (\bibinfo {year} {2016})}\BibitemShut
  {NoStop}%
\bibitem [{\citenamefont {Yang}\ \emph {et~al.}(2020)\citenamefont {Yang},
  \citenamefont {Duan}, \citenamefont {Wang},\ and\ \citenamefont
  {Jiang}}]{Yunjiang}%
  \BibitemOpen
  \bibfield  {author} {\bibinfo {author} {\bibfnamefont {J.}~\bibnamefont
  {Yang}}, \bibinfo {author} {\bibfnamefont {J.}~\bibnamefont {Duan}}, \bibinfo
  {author} {\bibfnamefont {Y.~J.}\ \bibnamefont {Wang}}, \ and\ \bibinfo
  {author} {\bibfnamefont {M.~Q.}\ \bibnamefont {Jiang}},\ }\href {\doibase
  10.1140/epje/i2020-11983-6} {\bibfield  {journal} {\bibinfo  {journal} {The
  European Physical Journal E}\ }\textbf {\bibinfo {volume} {43}},\ \bibinfo
  {pages} {56} (\bibinfo {year} {2020})}\BibitemShut {NoStop}%
\bibitem [{\citenamefont {Kleman}\ and\ \citenamefont
  {Lavrentovich}(2003)}]{Kleman}%
  \BibitemOpen
  \bibfield  {author} {\bibinfo {author} {\bibfnamefont {M.}~\bibnamefont
  {Kleman}}\ and\ \bibinfo {author} {\bibfnamefont {O.~D.}\ \bibnamefont
  {Lavrentovich}},\ }\href@noop {} {\emph {\bibinfo {title} {Soft Matter
  Physics: An introduction}}}\ (\bibinfo  {publisher} {Springer, New York},\
  \bibinfo {year} {2003})\BibitemShut {NoStop}%
\bibitem [{\citenamefont {Zaccone}(2013)}]{doi:10.1142/S0217984913300020}%
  \BibitemOpen
  \bibfield  {author} {\bibinfo {author} {\bibfnamefont {A.}~\bibnamefont
  {Zaccone}},\ }\href {\doibase 10.1142/S0217984913300020} {\bibfield
  {journal} {\bibinfo  {journal} {Modern Physics Letters B}\ }\textbf {\bibinfo
  {volume} {27}},\ \bibinfo {pages} {1330002} (\bibinfo {year} {2013})},\
  \Eprint {http://arxiv.org/abs/https://doi.org/10.1142/S0217984913300020}
  {https://doi.org/10.1142/S0217984913300020} \BibitemShut {NoStop}%
\bibitem [{\citenamefont {Laurati}\ \emph {et~al.}(2017)\citenamefont
  {Laurati}, \citenamefont {Ma\ss{}hoff}, \citenamefont {Mutch}, \citenamefont
  {Egelhaaf},\ and\ \citenamefont {Zaccone}}]{Laurati}%
  \BibitemOpen
  \bibfield  {author} {\bibinfo {author} {\bibfnamefont {M.}~\bibnamefont
  {Laurati}}, \bibinfo {author} {\bibfnamefont {P.}~\bibnamefont
  {Ma\ss{}hoff}}, \bibinfo {author} {\bibfnamefont {K.~J.}\ \bibnamefont
  {Mutch}}, \bibinfo {author} {\bibfnamefont {S.~U.}\ \bibnamefont {Egelhaaf}},
  \ and\ \bibinfo {author} {\bibfnamefont {A.}~\bibnamefont {Zaccone}},\ }\href
  {\doibase 10.1103/PhysRevLett.118.018002} {\bibfield  {journal} {\bibinfo
  {journal} {Phys. Rev. Lett.}\ }\textbf {\bibinfo {volume} {118}},\ \bibinfo
  {pages} {018002} (\bibinfo {year} {2017})}\BibitemShut {NoStop}%
\bibitem [{\citenamefont {Galloway}\ \emph {et~al.}(2020)\citenamefont
  {Galloway}, \citenamefont {Jerolmack},\ and\ \citenamefont
  {Arratia}}]{Arratia}%
  \BibitemOpen
  \bibfield  {author} {\bibinfo {author} {\bibfnamefont {K.~L.}\ \bibnamefont
  {Galloway}}, \bibinfo {author} {\bibfnamefont {D.~J.}\ \bibnamefont
  {Jerolmack}}, \ and\ \bibinfo {author} {\bibfnamefont {P.~E.}\ \bibnamefont
  {Arratia}},\ }\href {\doibase 10.1039/C9SM02482D} {\bibfield  {journal}
  {\bibinfo  {journal} {Soft Matter}\ }\textbf {\bibinfo {volume} {16}},\
  \bibinfo {pages} {4373} (\bibinfo {year} {2020})}\BibitemShut {NoStop}%
\bibitem [{\citenamefont {Schmid}\ and\ \citenamefont
  {Boas}(1935)}]{Schmid1935}%
  \BibitemOpen
  \bibfield  {author} {\bibinfo {author} {\bibfnamefont {E.}~\bibnamefont
  {Schmid}}\ and\ \bibinfo {author} {\bibfnamefont {W.}~\bibnamefont {Boas}},\
  }\enquote {\bibinfo {title} {Theorien der kristallplastizit{\"a}t und
  -festigkeit},}\ in\ \href {\doibase 10.1007/978-3-662-34532-0_9} {\emph
  {\bibinfo {booktitle} {Kristallplastizit{\"a}t: Mit Besonderer
  Ber{\"u}cksichtigung der Metalle}}}\ (\bibinfo  {publisher} {Springer Berlin
  Heidelberg},\ \bibinfo {address} {Berlin, Heidelberg},\ \bibinfo {year}
  {1935})\ pp.\ \bibinfo {pages} {279--301}\BibitemShut {NoStop}%
\bibitem [{\citenamefont {Courtney}(2005)}]{Courtney}%
  \BibitemOpen
  \bibfield  {author} {\bibinfo {author} {\bibfnamefont {T.~H.}\ \bibnamefont
  {Courtney}},\ }\href@noop {} {\emph {\bibinfo {title} {Mechanical Behavior of
  Materials}}}\ (\bibinfo  {publisher} {Waveland Press, Long Grove, IL},\
  \bibinfo {year} {2005})\BibitemShut {NoStop}%
\bibitem [{\citenamefont {Benzi}\ \emph {et~al.}(2016)\citenamefont {Benzi},
  \citenamefont {Sbragaglia}, \citenamefont {Bernaschi}, \citenamefont
  {Succi},\ and\ \citenamefont {Toschi}}]{Benzi}%
  \BibitemOpen
  \bibfield  {author} {\bibinfo {author} {\bibfnamefont {R.}~\bibnamefont
  {Benzi}}, \bibinfo {author} {\bibfnamefont {M.}~\bibnamefont {Sbragaglia}},
  \bibinfo {author} {\bibfnamefont {M.}~\bibnamefont {Bernaschi}}, \bibinfo
  {author} {\bibfnamefont {S.}~\bibnamefont {Succi}}, \ and\ \bibinfo {author}
  {\bibfnamefont {F.}~\bibnamefont {Toschi}},\ }\href {\doibase
  10.1039/C5SM01862E} {\bibfield  {journal} {\bibinfo  {journal} {Soft Matter}\
  }\textbf {\bibinfo {volume} {12}},\ \bibinfo {pages} {514} (\bibinfo {year}
  {2016})}\BibitemShut {NoStop}%
\bibitem [{\citenamefont {Milkus}\ and\ \citenamefont
  {Zaccone}(2016)}]{Milkus}%
  \BibitemOpen
  \bibfield  {author} {\bibinfo {author} {\bibfnamefont {R.}~\bibnamefont
  {Milkus}}\ and\ \bibinfo {author} {\bibfnamefont {A.}~\bibnamefont
  {Zaccone}},\ }\href {\doibase 10.1103/PhysRevB.93.094204} {\bibfield
  {journal} {\bibinfo  {journal} {Phys. Rev. B}\ }\textbf {\bibinfo {volume}
  {93}},\ \bibinfo {pages} {094204} (\bibinfo {year} {2016})}\BibitemShut
  {NoStop}%
\bibitem [{\citenamefont {Kelchner}\ \emph {et~al.}(1998)\citenamefont
  {Kelchner}, \citenamefont {Plimpton},\ and\ \citenamefont
  {Hamilton}}]{Plimpton}%
  \BibitemOpen
  \bibfield  {author} {\bibinfo {author} {\bibfnamefont {C.~L.}\ \bibnamefont
  {Kelchner}}, \bibinfo {author} {\bibfnamefont {S.~J.}\ \bibnamefont
  {Plimpton}}, \ and\ \bibinfo {author} {\bibfnamefont {J.~C.}\ \bibnamefont
  {Hamilton}},\ }\href {\doibase 10.1103/PhysRevB.58.11085} {\bibfield
  {journal} {\bibinfo  {journal} {Phys. Rev. B}\ }\textbf {\bibinfo {volume}
  {58}},\ \bibinfo {pages} {11085} (\bibinfo {year} {1998})}\BibitemShut
  {NoStop}%
\bibitem [{\citenamefont {Denisov}\ \emph {et~al.}(2015)\citenamefont
  {Denisov}, \citenamefont {Dang}, \citenamefont {Struth}, \citenamefont
  {Zaccone}, \citenamefont {Wegdam},\ and\ \citenamefont {Schall}}]{Schall}%
  \BibitemOpen
  \bibfield  {author} {\bibinfo {author} {\bibfnamefont {D.~V.}\ \bibnamefont
  {Denisov}}, \bibinfo {author} {\bibfnamefont {M.~T.}\ \bibnamefont {Dang}},
  \bibinfo {author} {\bibfnamefont {B.}~\bibnamefont {Struth}}, \bibinfo
  {author} {\bibfnamefont {A.}~\bibnamefont {Zaccone}}, \bibinfo {author}
  {\bibfnamefont {G.~H.}\ \bibnamefont {Wegdam}}, \ and\ \bibinfo {author}
  {\bibfnamefont {P.}~\bibnamefont {Schall}},\ }\href {\doibase
  10.1038/srep14359} {\bibfield  {journal} {\bibinfo  {journal} {Scientific
  Reports}\ }\textbf {\bibinfo {volume} {5}},\ \bibinfo {pages} {14359}
  (\bibinfo {year} {2015})}\BibitemShut {NoStop}%
\bibitem [{\citenamefont {Cubuk}\ \emph {et~al.}(2017)\citenamefont {Cubuk},
  \citenamefont {Ivancic}, \citenamefont {Schoenholz}, \citenamefont
  {Strickland}, \citenamefont {Basu}, \citenamefont {Davidson}, \citenamefont
  {Fontaine}, \citenamefont {Hor}, \citenamefont {Huang}, \citenamefont
  {Jiang}, \citenamefont {Keim}, \citenamefont {Koshigan}, \citenamefont
  {Lefever}, \citenamefont {Liu}, \citenamefont {Ma}, \citenamefont
  {Magagnosc}, \citenamefont {Morrow}, \citenamefont {Ortiz}, \citenamefont
  {Rieser}, \citenamefont {Shavit}, \citenamefont {Still}, \citenamefont {Xu},
  \citenamefont {Zhang}, \citenamefont {Nordstrom}, \citenamefont {Arratia},
  \citenamefont {Carpick}, \citenamefont {Durian}, \citenamefont {Fakhraai},
  \citenamefont {Jerolmack}, \citenamefont {Lee}, \citenamefont {Li},
  \citenamefont {Riggleman}, \citenamefont {Turner}, \citenamefont {Yodh},
  \citenamefont {Gianola},\ and\ \citenamefont {Liu}}]{Liu}%
  \BibitemOpen
  \bibfield  {author} {\bibinfo {author} {\bibfnamefont {E.~D.}\ \bibnamefont
  {Cubuk}}, \bibinfo {author} {\bibfnamefont {R.~J.~S.}\ \bibnamefont
  {Ivancic}}, \bibinfo {author} {\bibfnamefont {S.~S.}\ \bibnamefont
  {Schoenholz}}, \bibinfo {author} {\bibfnamefont {D.~J.}\ \bibnamefont
  {Strickland}}, \bibinfo {author} {\bibfnamefont {A.}~\bibnamefont {Basu}},
  \bibinfo {author} {\bibfnamefont {Z.~S.}\ \bibnamefont {Davidson}}, \bibinfo
  {author} {\bibfnamefont {J.}~\bibnamefont {Fontaine}}, \bibinfo {author}
  {\bibfnamefont {J.~L.}\ \bibnamefont {Hor}}, \bibinfo {author} {\bibfnamefont
  {Y.-R.}\ \bibnamefont {Huang}}, \bibinfo {author} {\bibfnamefont
  {Y.}~\bibnamefont {Jiang}}, \bibinfo {author} {\bibfnamefont {N.~C.}\
  \bibnamefont {Keim}}, \bibinfo {author} {\bibfnamefont {K.~D.}\ \bibnamefont
  {Koshigan}}, \bibinfo {author} {\bibfnamefont {J.~A.}\ \bibnamefont
  {Lefever}}, \bibinfo {author} {\bibfnamefont {T.}~\bibnamefont {Liu}},
  \bibinfo {author} {\bibfnamefont {X.-G.}\ \bibnamefont {Ma}}, \bibinfo
  {author} {\bibfnamefont {D.~J.}\ \bibnamefont {Magagnosc}}, \bibinfo {author}
  {\bibfnamefont {E.}~\bibnamefont {Morrow}}, \bibinfo {author} {\bibfnamefont
  {C.~P.}\ \bibnamefont {Ortiz}}, \bibinfo {author} {\bibfnamefont {J.~M.}\
  \bibnamefont {Rieser}}, \bibinfo {author} {\bibfnamefont {A.}~\bibnamefont
  {Shavit}}, \bibinfo {author} {\bibfnamefont {T.}~\bibnamefont {Still}},
  \bibinfo {author} {\bibfnamefont {Y.}~\bibnamefont {Xu}}, \bibinfo {author}
  {\bibfnamefont {Y.}~\bibnamefont {Zhang}}, \bibinfo {author} {\bibfnamefont
  {K.~N.}\ \bibnamefont {Nordstrom}}, \bibinfo {author} {\bibfnamefont {P.~E.}\
  \bibnamefont {Arratia}}, \bibinfo {author} {\bibfnamefont {R.~W.}\
  \bibnamefont {Carpick}}, \bibinfo {author} {\bibfnamefont {D.~J.}\
  \bibnamefont {Durian}}, \bibinfo {author} {\bibfnamefont {Z.}~\bibnamefont
  {Fakhraai}}, \bibinfo {author} {\bibfnamefont {D.~J.}\ \bibnamefont
  {Jerolmack}}, \bibinfo {author} {\bibfnamefont {D.}~\bibnamefont {Lee}},
  \bibinfo {author} {\bibfnamefont {J.}~\bibnamefont {Li}}, \bibinfo {author}
  {\bibfnamefont {R.}~\bibnamefont {Riggleman}}, \bibinfo {author}
  {\bibfnamefont {K.~T.}\ \bibnamefont {Turner}}, \bibinfo {author}
  {\bibfnamefont {A.~G.}\ \bibnamefont {Yodh}}, \bibinfo {author}
  {\bibfnamefont {D.~S.}\ \bibnamefont {Gianola}}, \ and\ \bibinfo {author}
  {\bibfnamefont {A.~J.}\ \bibnamefont {Liu}},\ }\href {\doibase
  10.1126/science.aai8830} {\bibfield  {journal} {\bibinfo  {journal}
  {Science}\ }\textbf {\bibinfo {volume} {358}},\ \bibinfo {pages} {1033}
  (\bibinfo {year} {2017})}\BibitemShut {NoStop}%
\bibitem [{\citenamefont {Richard}\ \emph {et~al.}(2020)\citenamefont
  {Richard}, \citenamefont {Ozawa}, \citenamefont {Patinet}, \citenamefont
  {Stanifer}, \citenamefont {Shang}, \citenamefont {Ridout}, \citenamefont
  {Xu}, \citenamefont {Zhang}, \citenamefont {Morse}, \citenamefont {Barrat},
  \citenamefont {Berthier}, \citenamefont {Falk}, \citenamefont {Guan},
  \citenamefont {Liu}, \citenamefont {Martens}, \citenamefont {Sastry},
  \citenamefont {Vandembroucq}, \citenamefont {Lerner},\ and\ \citenamefont
  {Manning}}]{Manning}%
  \BibitemOpen
  \bibfield  {author} {\bibinfo {author} {\bibfnamefont {D.}~\bibnamefont
  {Richard}}, \bibinfo {author} {\bibfnamefont {M.}~\bibnamefont {Ozawa}},
  \bibinfo {author} {\bibfnamefont {S.}~\bibnamefont {Patinet}}, \bibinfo
  {author} {\bibfnamefont {E.}~\bibnamefont {Stanifer}}, \bibinfo {author}
  {\bibfnamefont {B.}~\bibnamefont {Shang}}, \bibinfo {author} {\bibfnamefont
  {S.~A.}\ \bibnamefont {Ridout}}, \bibinfo {author} {\bibfnamefont
  {B.}~\bibnamefont {Xu}}, \bibinfo {author} {\bibfnamefont {G.}~\bibnamefont
  {Zhang}}, \bibinfo {author} {\bibfnamefont {P.~K.}\ \bibnamefont {Morse}},
  \bibinfo {author} {\bibfnamefont {J.-L.}\ \bibnamefont {Barrat}}, \bibinfo
  {author} {\bibfnamefont {L.}~\bibnamefont {Berthier}}, \bibinfo {author}
  {\bibfnamefont {M.~L.}\ \bibnamefont {Falk}}, \bibinfo {author}
  {\bibfnamefont {P.}~\bibnamefont {Guan}}, \bibinfo {author} {\bibfnamefont
  {A.~J.}\ \bibnamefont {Liu}}, \bibinfo {author} {\bibfnamefont
  {K.}~\bibnamefont {Martens}}, \bibinfo {author} {\bibfnamefont
  {S.}~\bibnamefont {Sastry}}, \bibinfo {author} {\bibfnamefont
  {D.}~\bibnamefont {Vandembroucq}}, \bibinfo {author} {\bibfnamefont
  {E.}~\bibnamefont {Lerner}}, \ and\ \bibinfo {author} {\bibfnamefont {M.~L.}\
  \bibnamefont {Manning}},\ }\href {\doibase 10.1103/PhysRevMaterials.4.113609}
  {\bibfield  {journal} {\bibinfo  {journal} {Phys. Rev. Materials}\ }\textbf
  {\bibinfo {volume} {4}},\ \bibinfo {pages} {113609} (\bibinfo {year}
  {2020})}\BibitemShut {NoStop}%
\bibitem [{\citenamefont {Ozawa}\ \emph {et~al.}(2018)\citenamefont {Ozawa},
  \citenamefont {Berthier}, \citenamefont {Biroli}, \citenamefont {Rosso},\
  and\ \citenamefont {Tarjus}}]{RossoPNAS}%
  \BibitemOpen
  \bibfield  {author} {\bibinfo {author} {\bibfnamefont {M.}~\bibnamefont
  {Ozawa}}, \bibinfo {author} {\bibfnamefont {L.}~\bibnamefont {Berthier}},
  \bibinfo {author} {\bibfnamefont {G.}~\bibnamefont {Biroli}}, \bibinfo
  {author} {\bibfnamefont {A.}~\bibnamefont {Rosso}}, \ and\ \bibinfo {author}
  {\bibfnamefont {G.}~\bibnamefont {Tarjus}},\ }\href {\doibase
  10.1073/pnas.1806156115} {\bibfield  {journal} {\bibinfo  {journal}
  {Proceedings of the National Academy of Sciences}\ }\textbf {\bibinfo
  {volume} {115}},\ \bibinfo {pages} {6656} (\bibinfo {year} {2018})},\ \Eprint
  {http://arxiv.org/abs/https://www.pnas.org/content/115/26/6656.full.pdf}
  {https://www.pnas.org/content/115/26/6656.full.pdf} \BibitemShut {NoStop}%
\bibitem [{\citenamefont {Grushin}(2020)}]{grushin2020topological}%
  \BibitemOpen
  \bibfield  {author} {\bibinfo {author} {\bibfnamefont {A.~G.}\ \bibnamefont
  {Grushin}},\ }\href@noop {} {\enquote {\bibinfo {title} {Topological phases
  of amorphous matter},}\ } (\bibinfo {year} {2020}),\ \Eprint
  {http://arxiv.org/abs/2010.02851} {arXiv:2010.02851 [cond-mat.dis-nn]}
  \BibitemShut {NoStop}%
\bibitem [{\citenamefont {Lema{\^i}tre}\ and\ \citenamefont
  {Maloney}(2006{\natexlab{b}})}]{Lemaitre2006}%
  \BibitemOpen
  \bibfield  {author} {\bibinfo {author} {\bibfnamefont {A.}~\bibnamefont
  {Lema{\^i}tre}}\ and\ \bibinfo {author} {\bibfnamefont {C.}~\bibnamefont
  {Maloney}},\ }\href@noop {} {\bibfield  {journal} {\bibinfo  {journal} {J.
  Stat. Phys.}\ }\textbf {\bibinfo {volume} {123}},\ \bibinfo {pages} {415}
  (\bibinfo {year} {2006}{\natexlab{b}})}\BibitemShut {NoStop}%
\end{thebibliography}%

\onecolumngrid
\appendix 
\clearpage

\section*{Supplementary Material}
In this Supplementary Material, we provide more details about the numerical simulations used in the main text, the computation of the Burgers vector and the validity of our results.

\subsection{Model and Simulation details}

We have used the Kremer-Grest model~\cite{Kremer1986,PhysRevB.102.024108}	of a coarse-grained polymer system consisting of linear chains of $50$ monomers. The polymer chain under consideration consisted of two different types of masses, where the two masses were chosen as $m_1=1$ and $m_2=3$. The geometry of the chain is such that the  masses are placed in alternating fashion. The total number of monomers in the system is $N=10000$.\\
\begin{figure}[ht]
\centering
\includegraphics[angle=-90,width=0.3\columnwidth]{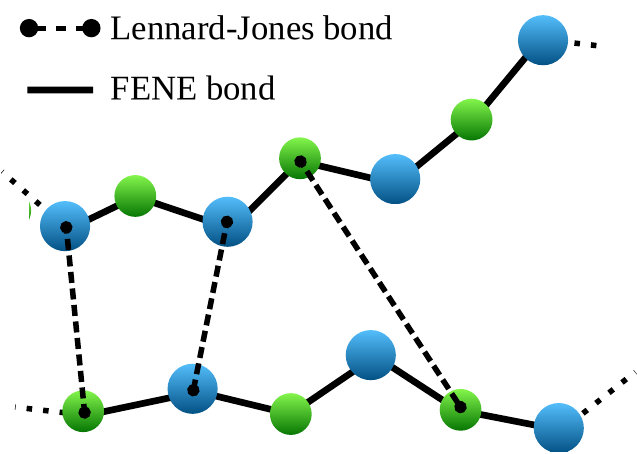}
 \caption{ Sketch of two alternating copolymer chains as they appear in the system. The Kremer-Grest model~\cite{Kremer1986} consisting of linear chains of 50 monomers is used. Some of the Lennard-Jones bonds between the chains are depicted as dashed lines. The FENE bonds along the polymer chain are represented as solid lines. The monomers with $m_1=1$ and $m_2=3$ alternate starting from the end of the chain (\textit{AB-configuration}).}. 
 \label{fig:model}
\end{figure}

\subsection*{Simulation of glass deformations}
The deformation have been performed with athermal quasi static (AQS) protocol \cite{Lemaitre2006} using the LAMMPS simulation package \cite{LAMMPS}. A glass sample initially quenched down to zero temperature is deformed by a quasi static shear procedure consisting in the relaxation of the system after each strain step ($\delta \gamma_{xz} = 0.001$). 

\begin{figure*}[ht]
    \centering
    \includegraphics[width=\linewidth]{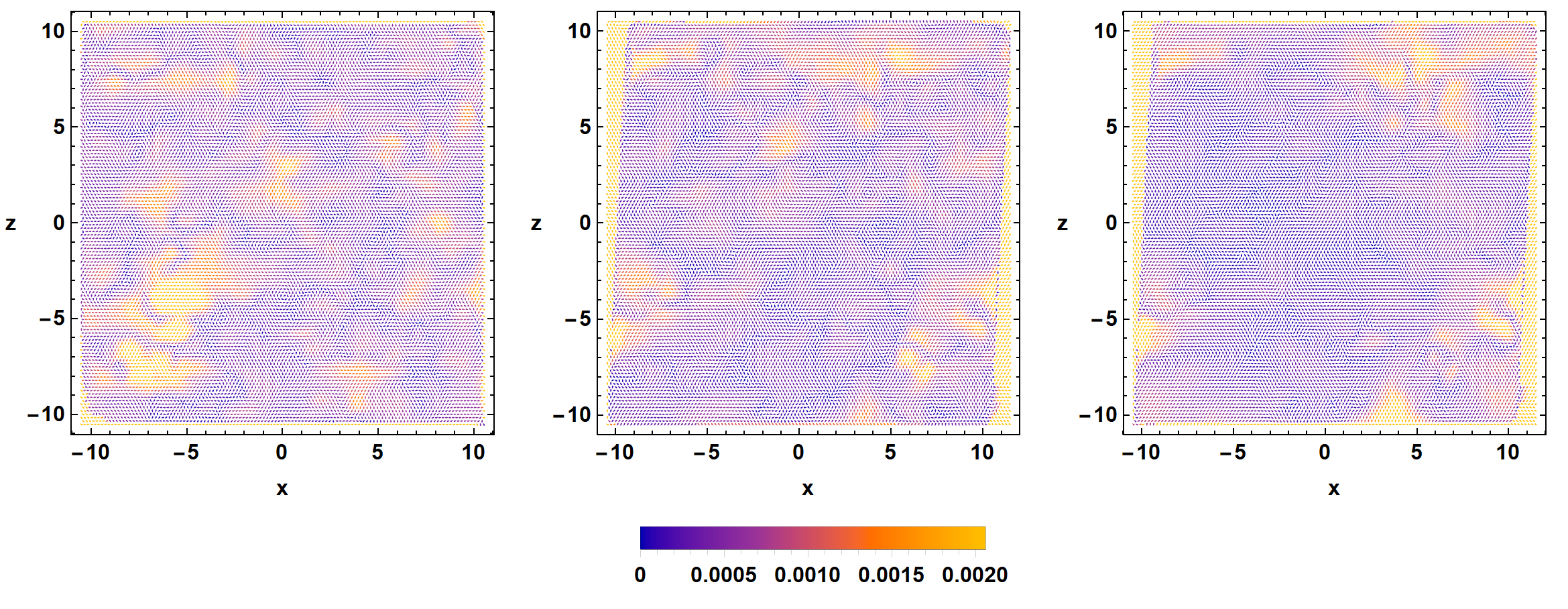}
    \caption{Three snapshots of the displacement configurations in the $(x,z)$ plane for different external strains $\gamma=0.01,0.05,0.06$. The colors indicate the amplitude of the vector field in each point as indicated in the legends.}
    \label{fig:app0}
\end{figure*}

\begin{figure}[ht]
    \centering
    \includegraphics[width=0.6\linewidth]{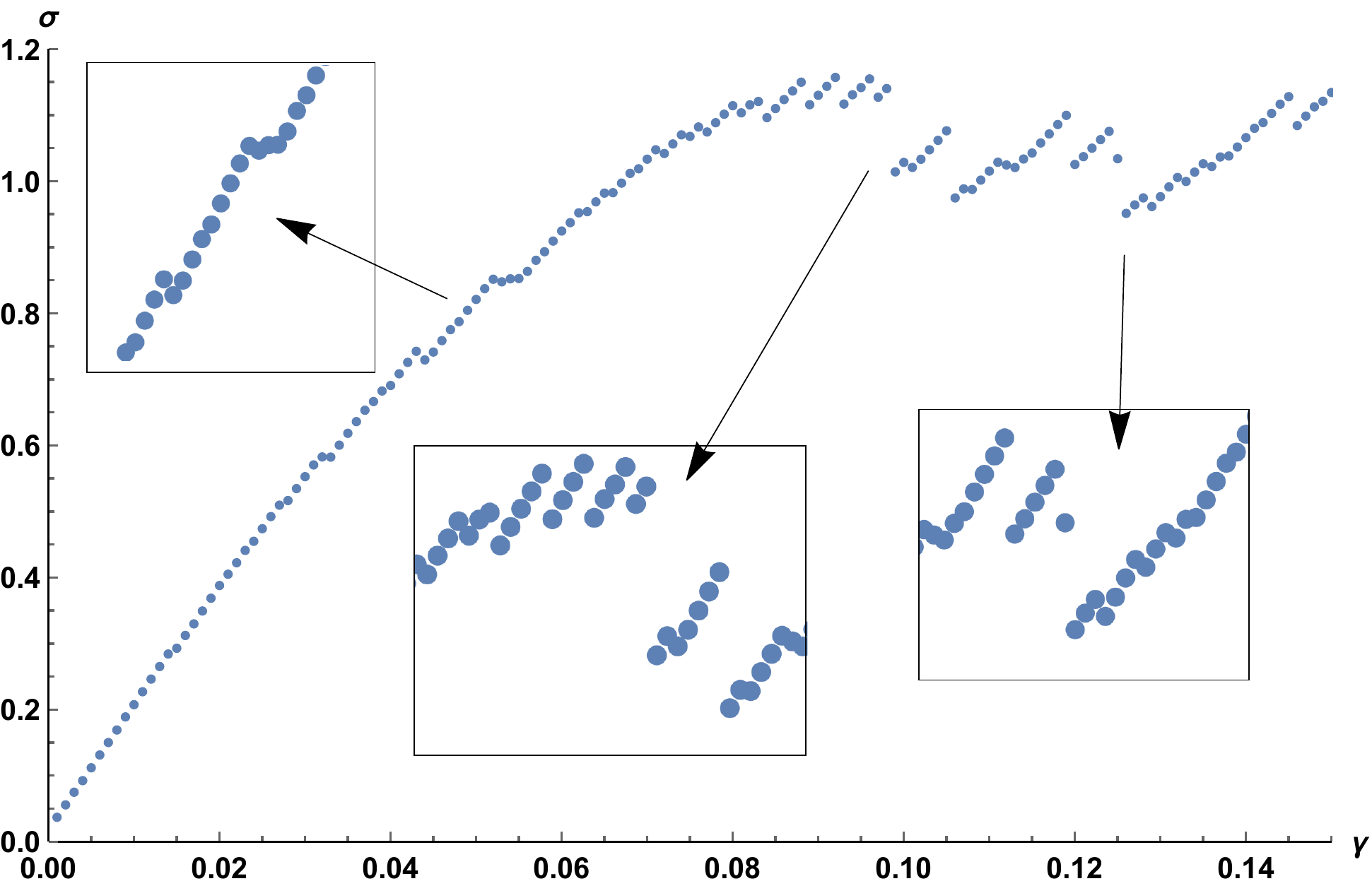}
    \caption{An example of the stress-strain curve with zooms on the most important plastic events.}
    \label{nn3}
\end{figure}

The snapshots of the system at each $\gamma$ are obtained after every relaxation along with stress information $\sigma$.
To obtain the non-affine displacement field $u^{\text{NA}}$ we use Equation 2 of the main article, where the total displacement $u$ is calculated from the difference of two subsequent snapshots and the affine part $u^{\text{A}}$ is defined by $\delta \gamma_{xz}$ ($x^{\text{A}}=z\, \delta\gamma_{xz}$). Snapshots of the displacements field are shown in Figure \ref{fig:app0}. An example of the stress-strain curve is presented in Figure \ref{nn3}.

\newpage
\subsection*{Determination of the Burgers vector}
\begin{figure}[ht]
    \centering
    \includegraphics[width=0.3\linewidth]{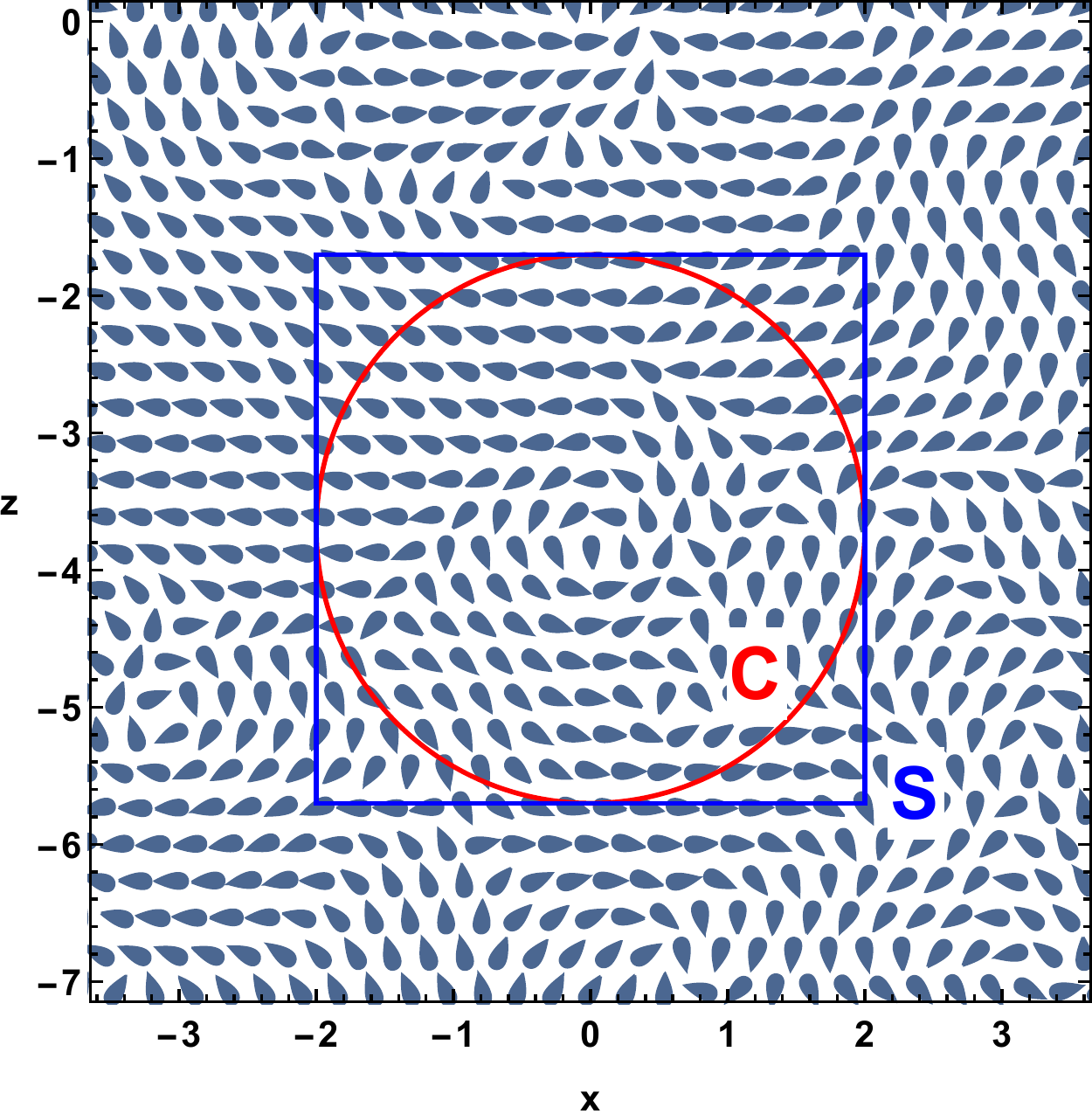}
    \qquad 
    \includegraphics[width=0.3\linewidth]{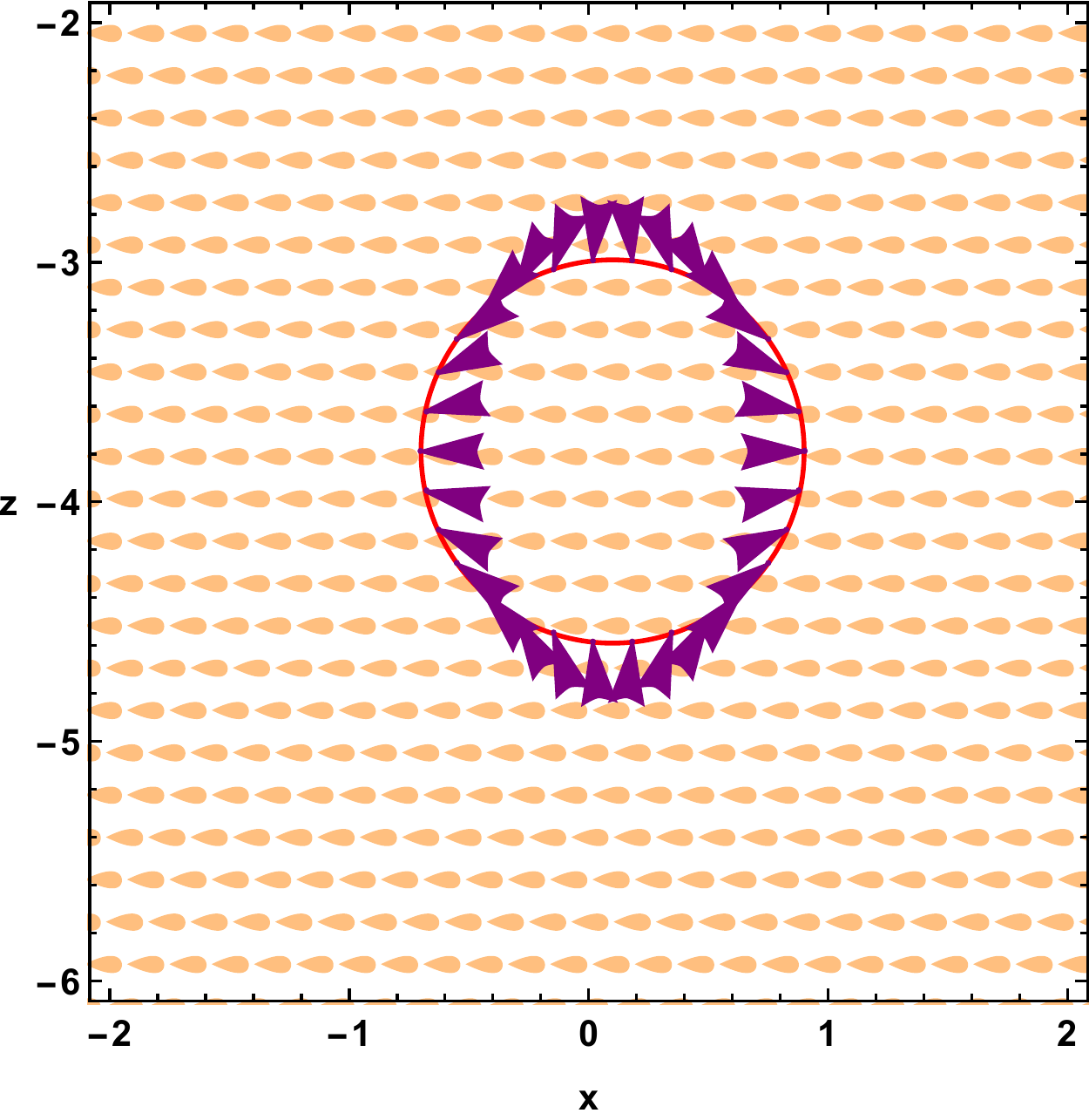}
    \qquad
     \includegraphics[width=0.3\linewidth]{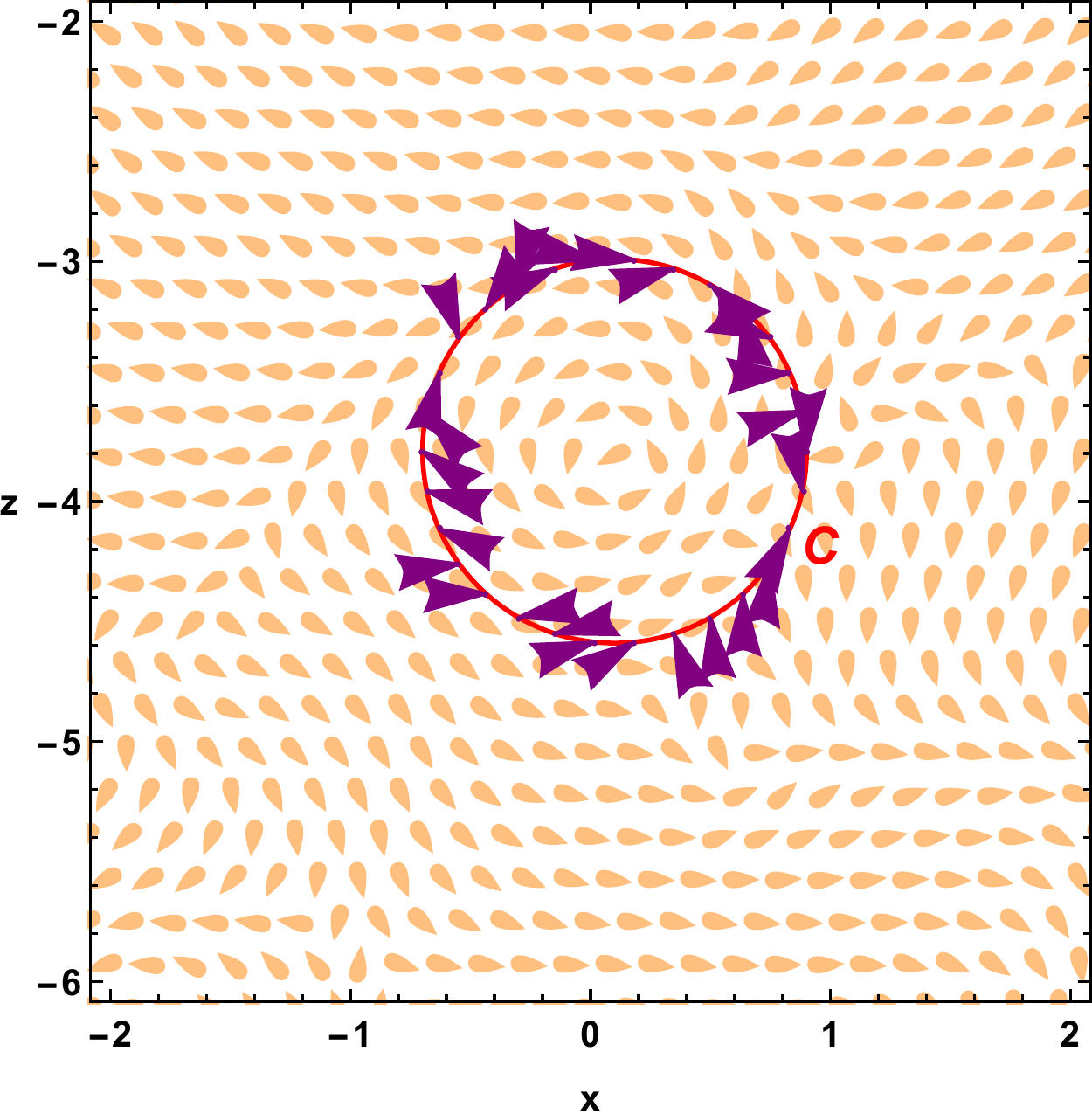}
    \caption{\textbf{Left: }The closed loops, circle $\mathcal{C}$ and square $\mathcal{S}$, considered in this work to compute the Burgers vector $b_i$. \textbf{Center: }The computation of the Burgers vector for the affine part of the displacement vectors. The partial integrands $\Delta u_i$ vanish in pairs leading to the expected result $b_i^{\text{aff}}=0$. \textbf{Right: }A graphical representation of the computation of the Burgers vector around a closed circle $\mathcal{C}$ for a displacement configuration with strain $\gamma=0.1$. The purple arrowheads indicate the partial integrands $\Delta u_i$.}
    \label{fig:app2}
\end{figure}

From the simulations, for every value of the external strain $\gamma$, we obtain the configuration of the displacement fields $\Vec{u}_(x,z)$ which are defined on a 2D $20 \times 20$ square box centered at the origin $(0,0)$. We have interpolated the discrete data using the built-in Mathematica "\textit{interpolation function}"\footnote{\url{https://reference.wolfram.com/language/ref/InterpolatingFunction.html}} together with the splines method and interpolating order $2$. We have checked explicitly (see more details below) that the results are not sensitive to this choice. From these data, we can compute numerically the corresponding strain tensor using
\begin{equation}
    \Vec{u}(x,z) \quad \Longrightarrow \quad \frac{du_i}{dx^k}(x,z)\,=\,\epsilon_{ik}(x,z)\,.
\end{equation}
Given a loop curve, parametrically described by a function $\Vec{x}(t)$, we can define the associated Burgers vector by computing the line integral of the vector field along the closed loop:
\begin{align}
    b_i\,\equiv\,-\,\oint_\mathcal{L}\,du_i\,=\,-\oint_\mathcal{L}\,\frac{du_i}{dx^k}\,dx^k\,=\,-\oint_\mathcal{L}\,\epsilon_{ik}\,dx^k\,=\,-\oint_{\mathcal{L}_t}\,\epsilon_{ik}\,\frac{dx^k}{dt}\,dt
\end{align}
For simplicity, we consider two different types of closed loops: a circle $\mathcal{C}$ of radius $R$ centered at $(x_c^{(0)},z_c^{(0)})$
\begin{equation}
\mathcal{C}:=\begin{cases}
&x_c(t)\,=\,x_c^{(0)}\,+\,R\,\cos{t}\\
&z_c(t)\,=\,z_c^{(0)}\,+\,R\,\sin{t}\\
& \text{with}\quad t\in\left[0,2\pi\right]
\end{cases}
\end{equation}
and a square $\mathcal{S}$ of side $2R$ centered at $(x_s^{(0)},z_s^{(0)})$
\begin{equation}
\mathcal{S}:=\begin{cases}
&l_1:\begin{cases}
&x_s(t)\,=\,x_s^{(0)}\,-\,R\,+\,t\\
&z_s(t)\,=\,z_s^{(0)}\,-\,R\,
\end{cases}\\[0.1cm]
&l_2:\begin{cases}
&x_s(t)\,=\,x_s^{(0)}\,+\,R\\
&z_s(t)\,=\,z_s^{(0)}\,-\,R\,+\,t
\end{cases}\\
&l_3:\begin{cases}
&x_s(t)\,=\,x_s^{(0)}\,+\,R\,-\,t\\
&z_s(t)\,=\,z_s^{(0)}\,+\,R\,
\end{cases}\\
&l_4:\begin{cases}
&x_s(t)\,=\,x_s^{(0)}\,-\,R\\
&z_s(t)\,=\,z_s^{(0)}\,+\,R\,-\,t
\end{cases}\\
&\text{with}\quad t\in\left[0,2\,R\right]
\end{cases}
\end{equation}
An example of the two closed curves is shown in the left panel of Fig.\ref{fig:app2}.

\begin{figure}[ht]
    \centering
    \includegraphics[width=0.35\linewidth]{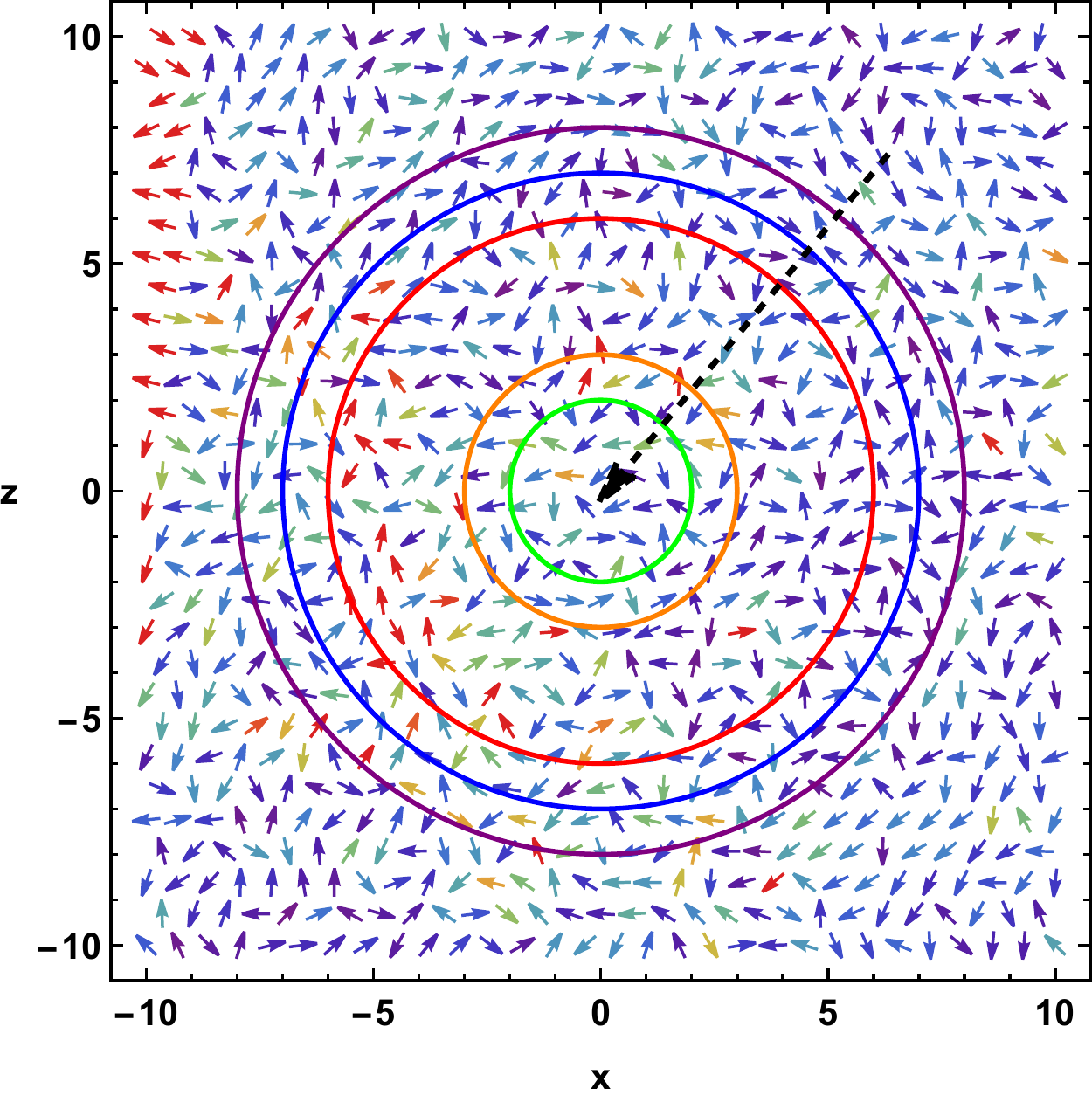}
    \quad 
     \includegraphics[width=0.62\linewidth]{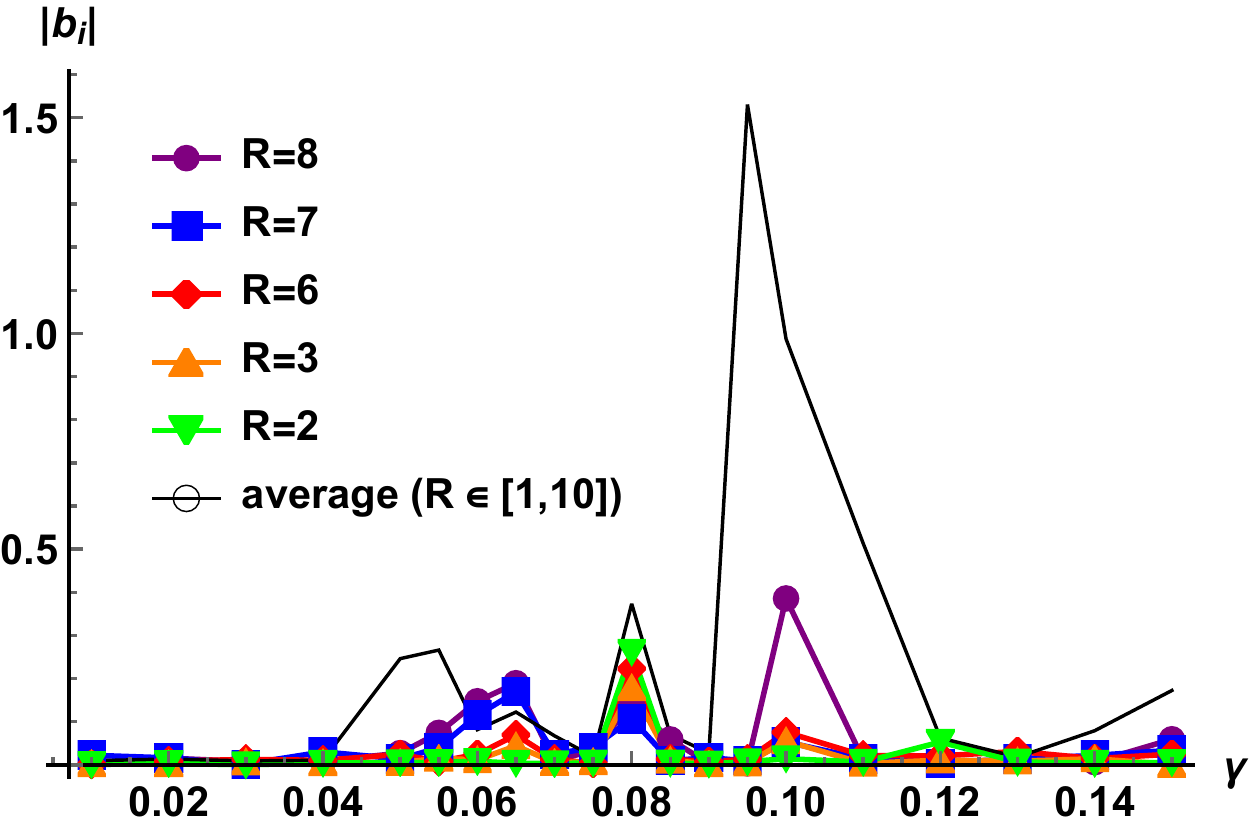}
     
     \vspace{1cm}
     
     \includegraphics[width=0.35\linewidth]{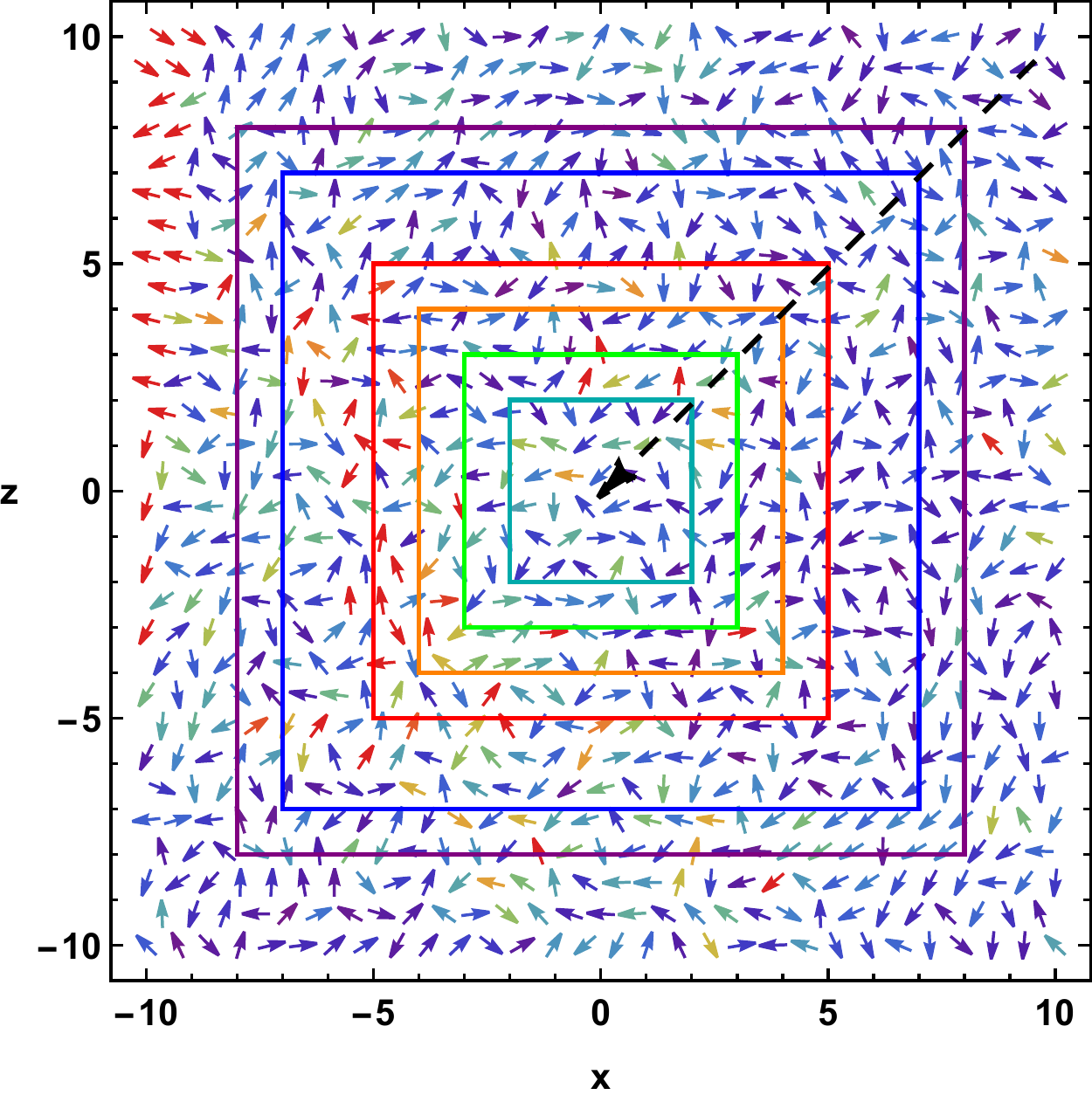}
    \quad 
     \includegraphics[width=0.62\linewidth]{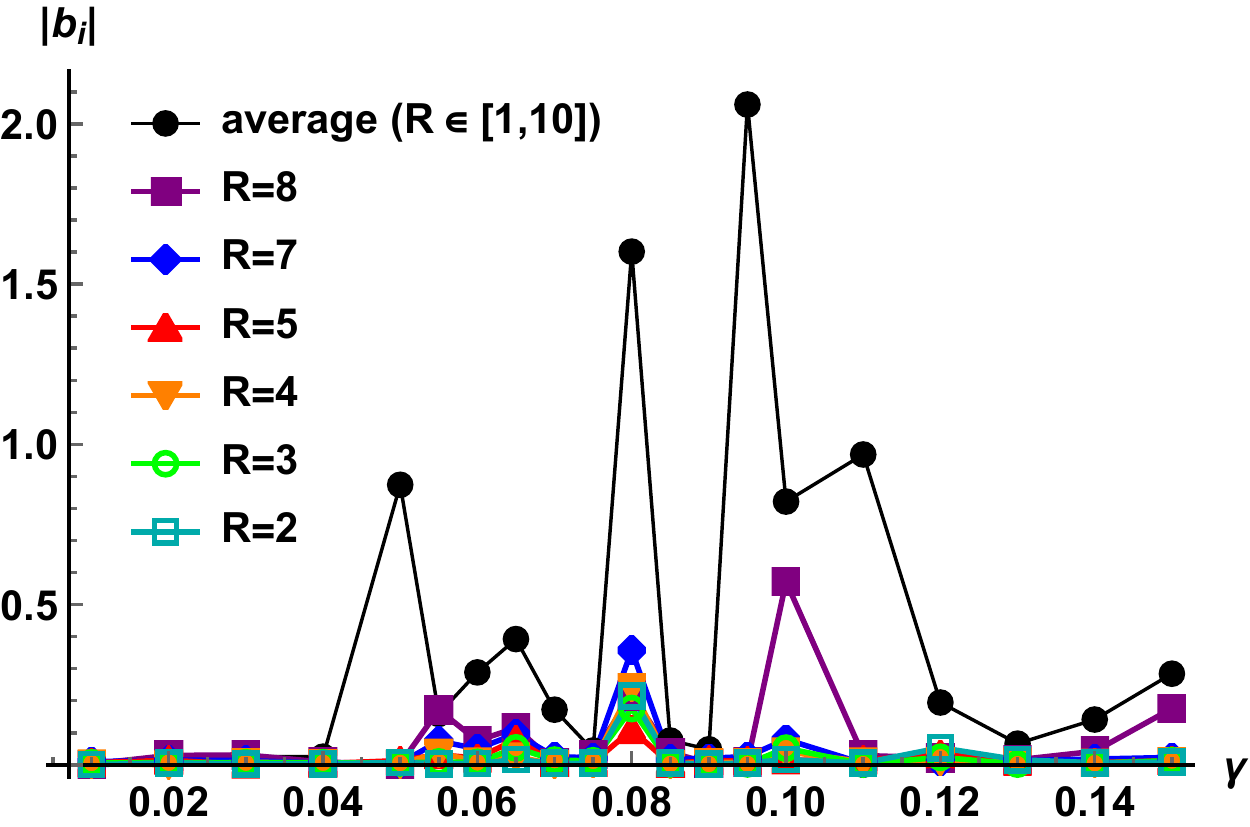}
     
     \vspace{0.4cm}
     
     \includegraphics[width=0.42\linewidth]{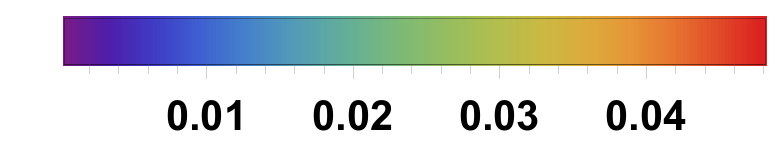}
    \caption{\textbf{Top: }The norm of the Burgers vector $|b_i|$ as a function of the external strain $\gamma$ for closed circles centered at the origin and with increasing radius $R \in \left[1,10\right]$. The yield point is at $\gamma \approx 0.1$. Two previous plastic events appear around $\gamma \approx 0.5,0.8$. \textbf{Bottom: }The norm of the Burgers vector $|b_i|$ as a function of the external strain $\gamma$ for closed squared centered at the origin and with increasing side size $R \in \left[1,10\right]$. The yield point is at $\gamma \approx 0.1$. Two previous plastic events appear around $\gamma \approx 0.5,0.8$. }
    \label{fig:app4}
\end{figure}

As a check of our routine, we have computed the Burgers vector for the affine part only of the displacements and we have always obtained zero as expected. An example of such check is shown in the right panel of Fig.\ref{fig:app2}. In the affine case, the partial integrands $\Delta u_i$ on the loop clearly vanish in pairs giving as a final result $0$.
\subsection*{Burgers loops size}
In the main text, we have shown only the computation of the norm of the Burgers vector $|b_i|$ for a single circular closed loop with radius $R=8$. In order to test in more detail the validity and robustness of our results, we have computed the norm of the Burgers vector in function of the external strain $\gamma$ for a large number of closed loops with different shapes. In particular, in Fig.\ref{fig:app4}, we show the results for a set of closed circles and squared centered at the origin and with decreasing radius and for their average. All the results robustly indicate the presence of two minor peaks located around $\gamma_1\approx 0.05$, $\gamma_2\approx 0/08$ corresponding to different plastic events before the yield point. Moreover, all the data displays a very strong peak around the yield point $\gamma_{\text{yield}}$.
\begin{figure}[ht]
    \centering
    \includegraphics[width=0.68\linewidth]{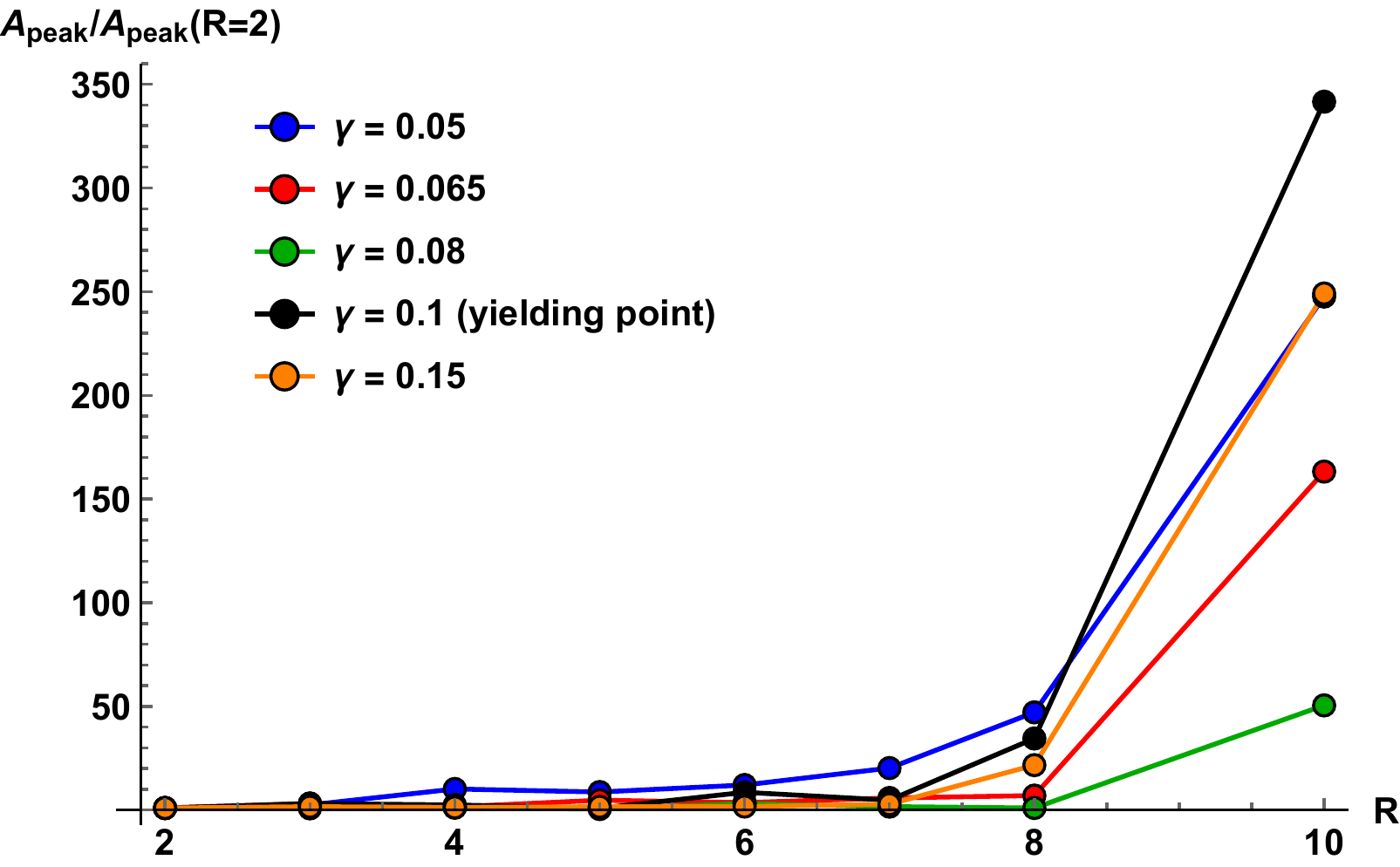}
    \caption{The normalized height of the Burgers vector peaks in function of the loop size $R$.}
    \label{nn2}
\end{figure}
We have studied the dependence of the amplitude of Burgers vector peaks upon the radius of the Burgers loop size systematically, results are shown in Fig.\ref{nn2} below. It is evident that the plastic event at $\gamma = 0.1$, which corresponds to the yielding transition, exhibits by far the strongest increase with the loop radius $R$. The increase for $\gamma = 0.1$ is indeed even larger than that for larger $\gamma$ values in the post-yielding flow regime. This observation fully supports the idea that the plastic event at yielding is a global event due to the system-spanning self-organization of a slip system, as discussed in the main article.
\subsection*{Topological invariant nature of the Burgers vector}
In Figure S.7 we show that the norm of the Burgers vector is independent (within numerical accuracy) of the shape of the loop or integration contour. This demonstrates the topological nature of the Burgers vector defined in the displacement field, i.e. that this is a topological invariant.
\begin{figure}[ht]
    \centering
    \includegraphics[width=0.68\linewidth]{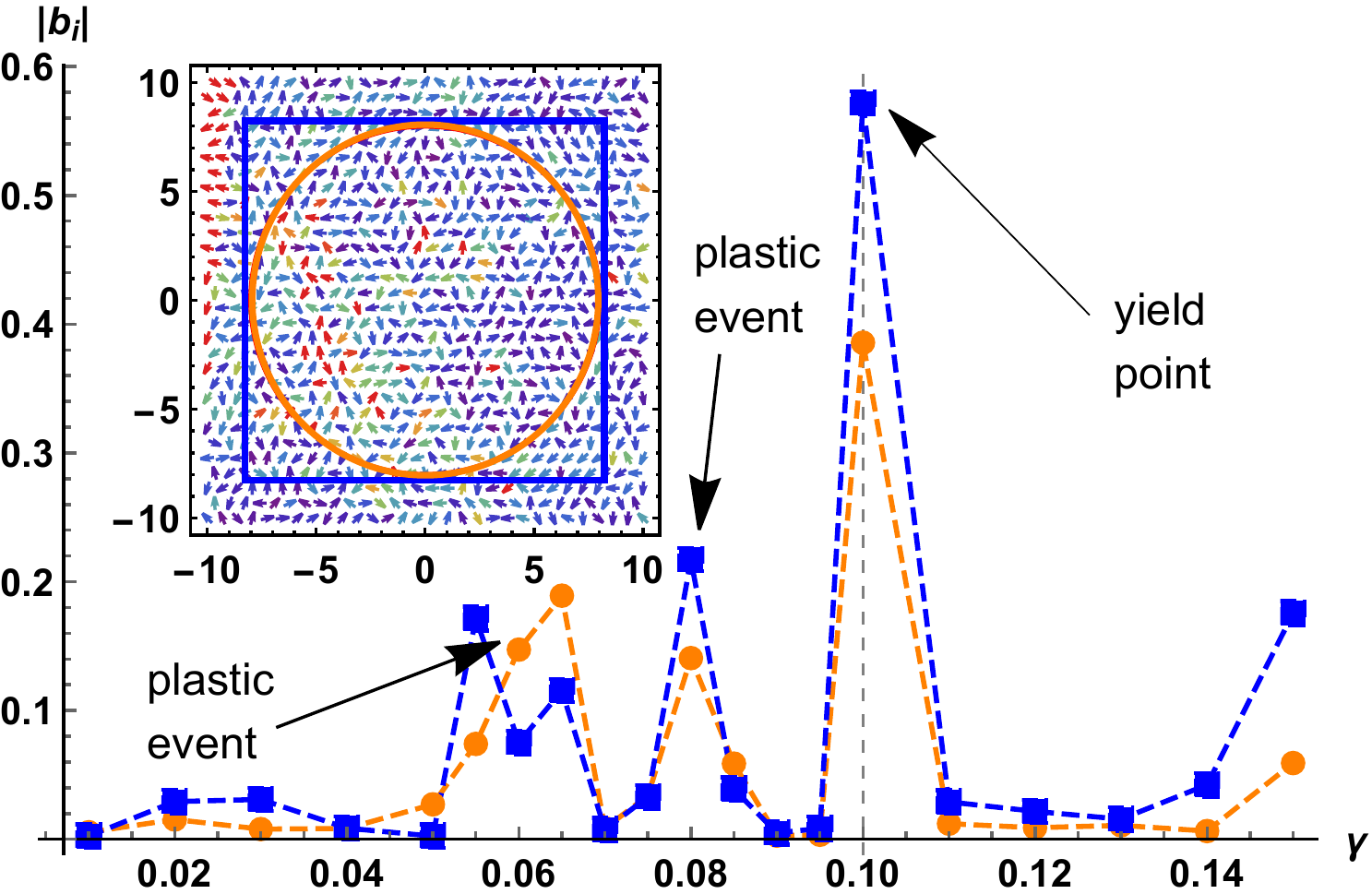}
    \caption{The norm of the Burgers vector $|b_i|$ obtained using circular (orange color) and square (blue color) closed loops.}
    \label{nn2}
\end{figure}
\subsection*{Analysis of the replica}
We performed 10 replicas of the glassy system. In Figure S.8 below we show 6 of these replicas. In each plot we also report the Burgers vector norm which features distinct sharp peaks in correspondence of each major plastic event. 
\begin{figure}
    \centering
    \includegraphics[width=\linewidth]{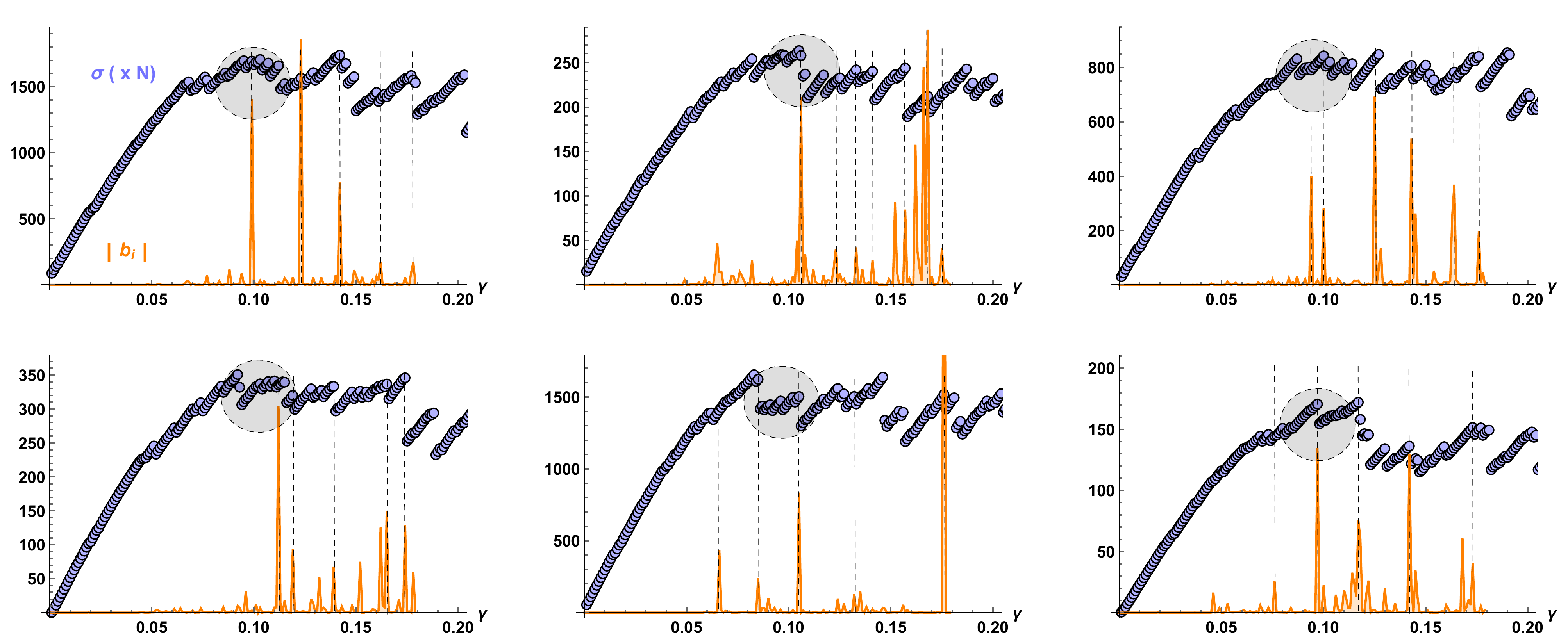}
 
    \caption{6 of the 10 replica analyzed. In purple the magnified stress and in orange the corresponding Burgers vector norm. The dashed lines guide the eyes of the reader to the position of the dominant peaks. The gray shaded area locates the yielding point.}
    \label{rep1}
\end{figure}

\subsection*{Interpolation method}
We have consistently used a quadratic spline interpolation in order to build a continuous displacement vector field out of the discrete simulations data. We have verified that our results do not depend on the type of interpolating functions (e.g. spline vs Hermite polynomials) and that they do not depend on the degree of the spline. As shown in the Figure S.9 below there is only a 4\% maximum variation in the Burgers vector norm across the different methods.
\begin{figure}[ht]
    \centering
    \includegraphics[width=0.6\linewidth]{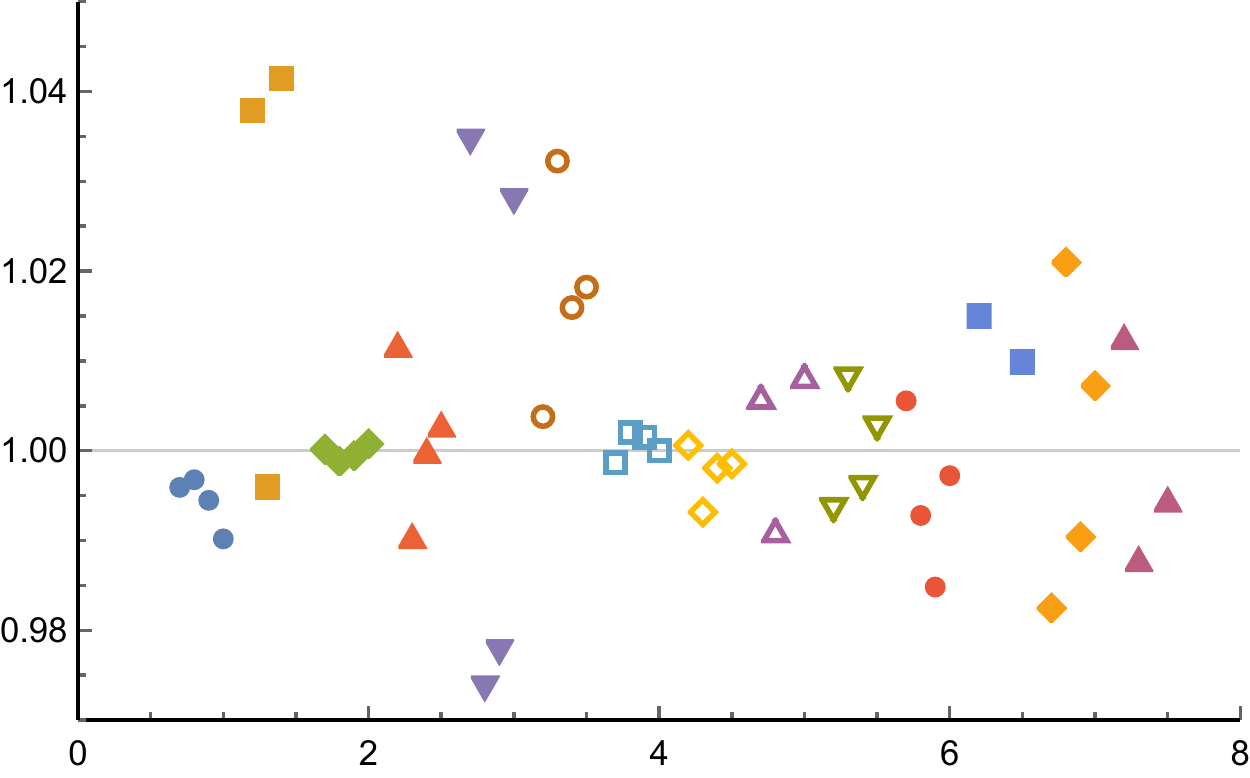}
    \caption{The values of the Burgers vector norm using different methods (splines and Hermite polynomials) and different interpolation orders normalized by the values used in the main text (quadratic splines). Different colors correspond to different Burgers loops. All the values lies in the range $\in [0.97,1.045]$ and they are therefore robust and independent of the interpolation procedure used.}
    \label{ee}
\end{figure}

\end{document}